\documentclass[longauth,bibyear]{aa} 
\usepackage[varg]{txfonts}

\usepackage{graphicx}
\usepackage{psfig,rotfloat,amsmath}
\usepackage{color}

\begin{document} 

\newcommand{\abs}[1]{\ensuremath{\lvert#1\rvert}}

   \title{
   Spectrophotometric properties of the nucleus of comet 67P/Churyumov-Gerasimenko from the OSIRIS instrument onboard the ROSETTA spacecraft.
}
   \subtitle{}

   \author{S. Fornasier
          \inst{1, 2}
           \and P. H. Hasselmann \inst{1,3}
           \and M.A. Barucci  \inst{1}
           \and C. Feller \inst{1, 2}
           \and S. Besse\inst{4}
           \and C. Leyrat\inst{1}
           \and L. Lara\inst{5}
           \and P. J. Gutierrez\inst{5}
           \and N. Oklay\inst{6}
           \and C. Tubiana\inst{6}
           \and F. Scholten \inst{7}
           \and H. Sierks\inst{6}
           \and C. Barbieri\inst{8}
           \and P. L. Lamy\inst{9}
           \and R. Rodrigo\inst{10,11}
           \and D. Koschny\inst{4}
           \and H. Rickman\inst{12,13}
           \and H. U. Keller\inst{14}
           \and J. Agarwal\inst{6}
           \and M. F. A'Hearn\inst{15}
J.-L. Bertaux\inst{16}\and
I. Bertini\inst{8}\and
G. Cremonese\inst{17}\and
V. Da Deppo\inst{18}\and
B. Davidsson\inst{12}\and
S. Debei\inst{19}\and
M. De Cecco\inst{20}\and
M. Fulle\inst{21}\and
O. Groussin\inst{9}\and
C. G\"uttler\inst{6}\and
S. F. Hviid\inst{7}\and
W. Ip\inst{22}\and
L. Jorda\inst{9}\and
J. Knollenberg\inst{7}\and
G. Kovacs\inst{6}\and
R. Kramm\inst{6}\and
E. K\"uhrt\inst{7}\and
M. K\"uppers\inst{23}\and
F. La Forgia\inst{8} \and
M. Lazzarin\inst{8} \and 
J. J. Lopez Moreno\inst{5}\and
F. Marzari\inst{8}\and
K.-D. Matz \inst{7} \and
H. Michalik\inst{24}\and
F. Moreno\inst{5}\and
S. Mottola\inst{7}\and
G. Naletto\inst{25, 26, 18}\and
M. Pajola\inst{26} \and 
A. Pommerol\inst{27} \and 
F. Preusker\inst{7} \and
X. Shi\inst{6}\and
C. Snodgrass\inst{6,28}\and
N. Thomas\inst{27}\and
J.-B. Vincent\inst{6}
 }

   \institute{LESIA, Observatoire de Paris, CNRS, UPMC Univ Paris 06, Univ. Paris-Diderot, 5 Place J. Janssen,  92195 Meudon Pricipal Cedex, France
              \email{sonia.fornasier@obspm.fr}
                \and Univ Paris Diderot, Sorbonne Paris Cit\'{e}, 4 rue Elsa Morante, 75205 Paris Cedex 13, France
                \and Observat\'orio Nacional, General Jos\'e Cristino 77, S\~ao Cristov\~ao, Rio de Janeiro, Brazil 
                \and Research and Scientific Support Department, European Space Agency, 2201 Noordwijk, The Netherlands 
                 \and Instituto de Astrof\'isica de Andaluc\'ia -- CSIC, 18080 Granada, Spain 
                 \and Max-Planck-Institut f\"ur Sonnensystemforschung, Justus-von-Liebig-Weg, 3 37077 G\"ottingen, Germany 
                 \and
                 Institute of Planetary Research, DLR, Rutherfordstrasse 2, 12489 Berlin, Germany 
                 \and
                Department of Physics and Astronomy "G. Galilei", University of Padova, Vic. Osservatorio 3, 35122 Padova, Italy
                \and
                Laboratoire d’Astrophysique de Marseille UMR 7326, CNRS \& Aix Marseille Universit\'e, 13388 Marseille Cedex 13, France
                \and
                Centro de Astrobiolog\'ia, CSIC-INTA, 28850 Torrej\'on de Ardoz, Madrid, Spain 
                \and
                International Space Science Institute, Hallerstrasse 6, 3012 Bern, Switzerland 
                \and
                Department of Physics and Astronomy, Uppsala University, 75120 Uppsala, Sweden 
                \and
                PAS Space Reserch Center, Bartycka 18A, 00716 Warszawa, Poland  
                \and
                Institute for Geophysics and Extraterrestrial Physics, TU Braunschweig, 38106 Braunschweig, Germany 
                \and
                Department for Astronomy, University of Maryland, College Park, MD 20742-2421, USA 
                \and
                LATMOS, CNRS/UVSQ/IPSL, 11 Boulevard d'Alembert, 78280 Guyancourt, France 
                \and
                INAF--Osservatorio Astronomico di Padova, Vicolo dell'Osservatorio 5, 35122 Padova, Italy 
                \and
                CNR--IFN UOS Padova LUXOR, Via Trasea 7, 35131 Padova, Italy 
                \and
                Department of Mechanical Engineering -- University of Padova, Via Venezia 1, 35131 Padova, Italy 
                \and
                UNITN, Universit\'a di Trento, Via Mesiano, 77, 38100 Trento, Italy 
                \and
                INAF -- Osservatorio Astronomico di Trieste, via Tiepolo 11, 34143 Trieste, Italy 
                \and
                Institute for Space Science, National Central University, 32054 Chung-Li, Taiwan
                \and
                ESA/ESAC, PO Box 78, 28691 Villanueva de la Ca\~nada, Spain 
                \and
                Institut f\"ur Datentechnik und Kommunikationsnetze, 38106 Braunschweig, Germany 
                \and
                Department of Information Engineering - University of Padova, Via Gradenigo 6, 35131 Padova, Italy 
                \and Center of Studies and Activities for Space (CISAS) "G. Colombo", University of Padova, Via Venezia 15, 35131 Padova, Italy
                 \and
                Physikalisches Institut, Sidlerstrasse 5, University of Bern, CH-3012 Bern, Switzerland 
                \and
                 Planetary and Space Sciences, Department of Physical Sciences, The Open University, Walton Hall, Milton Keynes, MK7 6AA, UK 
            }

   \date{Received Feb. 2015, ....}

\newpage

 
  \abstract
    {The Rosetta mission of the European
Space Agency has been orbiting the comet 67P/Churyumov-Gerasimenko
(67P) since August 2014 and is now in its escort phase. A large complement of scientific experiments designed to complete the most detailed study of a comet ever attempted are onboard Rosetta.}
   {We present results for the photometric and spectrophotometric properties of the nucleus of 67P derived from the OSIRIS imaging system, which consists of a Wide Angle Camera (WAC) and a Narrow Angle Camera (NAC). The observations presented here were performed during  July  and the beginning of August 2014, during the approach phase, when OSIRIS was mapping the surface of the comet with several filters  at different phase angles (1.3$^{\circ}$--54$^{\circ}$).
The resolution reached up to 2.1 m/px. }
   {The OSIRIS images were processed with the OSIRIS standard pipeline, then converted into $I/F$ radiance factors and corrected for the illumination conditions at each pixel using the Lommel-Seeliger disk law. Color cubes of the surface were produced by stacking registered and illumination-corrected images. Furthermore, photometric analysis was performed both on disk-averaged photometry in several filters and on disk-resolved images acquired with the NAC orange filter, centered at 649 nm, using Hapke modeling.}
   {The disk-averaged phase function of the nucleus of 67P shows a strong opposition surge with a G parameter value of  -0.13$\pm$0.01 in the HG system
formalism and an absolute magnitude $H_v(1,1,0)$ = 15.74$\pm$0.02 mag. The integrated spectrophotometry in 20 filters covering the 250-1000 nm wavelength range shows a red spectral behavior, without clear absorption bands except for a potential absorption centered at $\sim$ 290 nm that is possibly due to SO$_2$ ice.  The nucleus shows strong phase reddening, with disk-averaged spectral slopes increasing from 11\%/(100 nm) to 16\%/(100 nm) in the 1.3$^{\circ}$--54$^{\circ}$ phase angle range. The geometric albedo of the comet is 6.5$\pm$0.2\% at 649 nm, with local variations of up to $\sim$ 16\% in the Hapi region. From the disk-resolved images we computed the spectral slope together with local spectrophotometry and identified three distinct groups of regions (blue, moderately red, and red). The Hapi region is the brightest, the bluest in term of spectral slope, and the most active surface on the comet. Local spectrophotometry shows an enhancement of the flux in the 700-750 nm that is associated with coma emissions.} 
   {}

\titlerunning{Physical properties of the 67P comet nucleus}
   \maketitle
%

\section{Introduction}

The Rosetta spacecraft arrived on 6 August 2014 at comet 67P/Churyumov-Gerasimenko
(67P) after ten years of interplanetary journey. The nucleus has been mapped by the Optical, Spectroscopic, and Infrared Remote Imaging System (OSIRIS) (Keller et al., 2007), which comprises a Narrow Angle Camera (NAC) for nucleus surface and dust studies, and a Wide Angle Camera (WAC) for the wide-field coma investigations. 
From the beginning of July, images obtained with different filters (from 240 to 1000 nm) with the NAC camera allowed us to obtain color mapping of the comet nucleus with an increasing spatial resolution during the global mapping phase at the comet. Observations  will continue during the nominal mission until the end of December 2015.\\ 
The very first results on the comet images and spectroscopy obtained with the OSIRIS and VIRTIS instruments (Sierks et al., 2015; Thomas et al., 2015;  Capaccioni et al., 2015) reveal that the
nucleus of  67P has surface characteristics that are very different, in terms of shape, complex morphology, and spectral properties, from those of the other cometary nuclei that were visited by space missions.
Before the ESA Rosetta interplanetary mission, exploration of comets was considered by different space agencies. These began in 1986 with the flyby of comet 1P/ Halley (Keller et al., 1986) by the ESA Giotto mission, launched in 1985 and followed by a flotilla of spacecraft from the Japanese and Russian space agencies. The Giotto Extended Mission (GEM)  made the second flyby of comet 26P/Grigg-Skjellerup in July 1992. In 2001, the NASA Deep Space 1 mission flew by comet 19P/Borrelly (Soderblom et al., 2002). The NASA Discovery program selected the missions Stardust and Deep Impact. The sample-return mission Stardust had flown by comet 81P/Wild 2 in 2004 (Brownlee et al., 2004), while in  July 2005, the mission Deep Impact could observe the collision induced by an impactor on the nucleus of comet 9P/Tempel 1 (A’Hearn et al, 2005).  The comet 103P/Hartley 2 was visited in November 2010 by the redirected Deep Impact mission,  named  EPOXI (A'Hearn et al., 2011), while Stardust, renamed NExT, was redirected to fly by comet  9P/Tempel 1 in February 2011  (Veverka et al., 2013).  These missions (see Barucci et al. (2011) for a complete review) yielded an incredible knowledge of comets, unveiling their surface structures and many other properties. The known cometary nuclei  are irregularly shaped, and each of them presents a variety of different morphologies (e.g., depressions, ridges, chains of hills, smooth areas, rough terrains, layers, and craters) and different spectral properties. 
The Rosetta mission is the first mission following and orbiting a comet from $\sim$ 4 AU inbound to 2 AU outbound including the perihelion passage at 1.24 AU. The OSIRIS imaging system is the first instrument with the capability of mapping a comet surface at such a high resolution (lower than 20 cm/px at best) with 20 filters and covering various phase angle and illumination conditions. 

 We here present  results for  the spectrophotometric properties
of the nucleus of 67P derived from the OSIRIS observations obtained from July to mid-August 2014, during the comet approach phase and the first bound
orbits.   We present the global and local analysis of the photometric properties of the nucleus and spectrophotometry.
 These data give indications for the properties of the comet nucleus and allow us to investigate the heterogeneity of the
nucleus at several scales, both in terms of albedo and composition.

  \section{Observations and data reduction}
  
The NAC  has a field of view of $2.2\ensuremath{^\circ}\times2.2 \ensuremath{^\circ}$ and was designed to obtain high-resolution images with different filters in the near-UV to near-IR range (240-1000 nm). These filters are optimized for mineralogical studies of the nucleus. The WAC  has a wide field of view ($11.6\ensuremath{^\circ}\times12.1 \ensuremath{^\circ}$) and a set of narrow-band filters devoted to study gaseous species in the coma. The NAC and WAC cameras have unobstructed mirror systems with
focal lengths of 72 cm and 13 cm, respectively. Both cameras are equipped with 2048$\times$2048 pixel
CCD detectors with a pixel size of 13.5 $\mu$m. The image scale is
3.9 "/px for the NAC and 20.5 "/px for the WAC. We refer to Keller et al. (2007) for a more detailed description of the OSIRIS cameras.

From mid-July onward, the comet was resolved,  and  OSIRIS has mapped the nucleus surface with increasing spatial resolution. We present the results of OSIRIS observations obtained from the very first resolved images up to the beginning of August, when the spacecraft began to orbit the comet at a distance of 110 km. In this time frame, OSIRIS has mapped the surface of the comet with several filters  at different phase angles (1.3$^{\circ}$--54$^{\circ}$) and with a resolution of 2.1 m/px (Table~\ref{observing}). 
 The data were reduced using the OSIRIS standard pipeline up to level 3, following the data reduction steps described in K\"uppers et al. (2007) and Tubiana et al. (2015a), which include correction for bias, flat field, geometric distortion, and calibration in absolute flux.
 The absolute calibration factors that convert digital units into $W m^{-2} nm^{-1}sr^{-1}$ (referred to the central wavelength of each filter) were recently revised in the OSIRIS pipeline and are slightly updated compared with
those used before the Rosetta hibernation; they were  successfully tested on Vesta (Fornasier et al., 2011).  The absolute calibration factors were derived from Vega observations acquired during the instrument calibration campaign on 18 May 2014, and computed using the Vega and the Sun flux standard spectra from the HST CALSPEC catalog ($www.caha.es/pedraz/SSS/HST\_CALSPEC$). 
 
 For each image, we computed the radiance factor (also known as $I/F$) for each
pixel,
\begin{equation}
Radiance Factor (\lambda)  = \frac{\pi I(i,e,\alpha,\lambda)}{F_{\lambda}},
\end{equation}
where I is the observed scattered radiance, $F_{\lambda}$ the incoming solar irradiance at the heliocentric distance of the comet, and  $i$, $e,$ and $\alpha$ are the incidence, emission, and phase angles, respectively.
The solar irradiance $F_{\lambda}$ is wavelength dependent and was calculated at the central wavelength
of each filter to be consistent with the methodology applied to derive the absolute calibration factors.

\begin{table*}
\begin{center}
\caption{Observing conditions for the OSIRIS data. The filter names are followed by their central wavelength in nm at the first time that they appear in the table. The time refers to the start time of the first image of each sequence. $\Delta$ is the distance between the Rosetta spacecraft and comet 67P.}
\small{
\label{observing}
\begin{tabular}{l l l c c c } 
Camera & Time &        filters                                               & phase ($^{\circ}$) & $\Delta$ (km) & res. (m/px)  \\ \hline
NAC & 2014-07-21T14.34.16  & F22 (649.2), F23 (535.7), F24 (480.7), F16 (360.0),   &  5.3   & 4952   &  93 \\
    &                     & F27 (701.2), F28 (743.7), F41 (882.1), F51 (805.3),               &    &              \\
    &                      &    F61 (931.9), F71 (989.3), F15 (269.3), \& WAC-F71 (325)           &          &                         \\  
NAC & 2014-07-21T15.50.16  & F22, F23, F24, F16, F27, F28, F41, F51, F61, F71, F15, \& WAC-F71                        &  5.3   & 4915    &  93 \\
NAC & 2014-07-21T17.06.16  & F22, F23, F24, F16, F27, F28, F41, F51, F61, F71, F15, \& WAC-F71                        &  5.2   & 4878    &  92 \\  
NAC & 2014-07-21T18.22.16  & F22, F23, F24, F16, F27, F28, F41, F51, F61, F71, F15, \& WAC-F71                        &  5.1   & 4842    &  91 \\  
NAC & 2014-07-21T19.38.16  & F22, F23, F24, F16, F27, F28, F41, F51, F61, F71, F15, \& WAC-F71                         &  5.0   & 4805    &  91 \\
NAC & 2014-07-21T20.54.16  & F22, F23, F24, F16, F27, F28, F41, F51, F61, F71, F15, \& WAC-F71                         &  5.0   & 4768    &  90 \\
NAC & 2014-07-21T22.10.16  & F22, F23, F24, F16, F27, F28, F41, F51, F61, F71, F15, \& WAC-F71                         &  4.9   & 4731    &  89 \\ 
NAC & 2014-07-21T23.26.16  & F22, F23, F24, F16, F27, F28, F41, F51, F61, F71, F15, \& WAC-F71                         &  4.8   & 4695    &  88 \\  
NAC & 2014-07-22T00.42.16  & F22, F23, F24, F16, F27, F28, F41, F51, F61, F71, F15, \& WAC-F71                         &  4.8   & 4658    &  88 \\  
NAC & 2014-07-25T08.11.21  & F22, F23, F24, F16, F28, F41, F71, \& WAC-F71                                          &  2.2   & 3092    &  58 \\
NAC & 2014-07-25T09.34.58  & F22, F23, F24, F16, F28, F41, F71, \& WAC-F71                                             &  2.2   & 3074    &  58 \\
NAC & 2014-07-25T11.00.58  & F22, F23, F24, F16, F28, F41, F71, \& WAC-F71                                             &  2.2   & 3055    &  58 \\ 
NAC & 2014-07-25T12.26.58  & F22, F23, F24, F16, F28, F41, F71, \& WAC-F71                                            &  2.2   & 3037    &  57 \\ 
NAC & 2014-07-25T13.52.58  & F22, F23, F24, F16, F28, F41, F71, \& WAC-F71                                             &  2.1   & 3019    &  57 \\
NAC & 2014-07-25T15.18.58  & F22, F23, F24, F16, F28, F41, F71, \& WAC-F71                                             &  2.1   & 3000    &  57 \\    
NAC & 2014-07-25T18.10.58  & F22, F23, F24, F16, F28, F41, F71, \& WAC-F71                                             &  2.0   & 2963    &  56 \\  
NAC & 2014-07-25T19.55.58  & F22, F23, F24, F16, F28, F41, F71, \& WAC-F71                                             &  2.0   & 2940    &  55 \\  
NAC & 2014-07-28T22.10.25  & F27, F28, F41, F51, F61, F71, \& WAC-F71                                               &    1.3 & 1984    &  37 \\
WAC & 2014-07-28T22.10.25  & F18 (613), F71 (325), F13 (375), F14 (388), F31 (246),                                        &    1.3 & 1984    &  195 \\
WAC & 2014-07-28T22.10.25  & F41 (259), F51 (295), F61 (309),  F81 (336)                                                  &         &     & \\         
NAC & 2014-07-28T23.25.31  & F22, F23, F24, F16, F27, F28, F41, F51, F61, F71, F15, \& WAC-F71                         &    1.3 & 1968    &  37 \\    
WAC & 2014-07-28T23.25.31  &  F18, F71, F13, F14, F31, F41, F51, F61, F81                                                &    1.3 & 1968    &  195 \\    
NAC & 2014-07-29T00.45.31  & F22, F23, F24, F16, F27, F28, F41, F51, F61, F71, F15                                        &    1.3 & 1951    &  37 \\    
WAC & 2014-07-29T00.45.31  & F18, F71, F13, F14, F31, F41, F51, F61, F81                                                &    1.3 & 1968    &  195 \\    
NAC & 2014-08-01T11.50.14  & F82 (649.2), F23, F24, F27, F28, F41, F71, \& WAC-F71                                     &    9.0 &  826    &  16 \\
NAC & 2014-08-01T13.20.42  & F82, F23, F24, F27, F28, F41, F71, \& WAC-F71                                             &   9.3  &  807    &  15 \\   
NAC & 2014-08-01T14.43.48  & F82, F23, F24, F27, F28, F41, F71, \& WAC-F71                                             &   9.7  &  790    &  15 \\   
NAC & 2014-08-01T16.08.14  & F82, F23, F24, F27, F28, F41, F71, \& WAC-F71                                             & 10.1   &  772    &  14 \\  
NAC & 2014-08-01T17.26.34  & F82, F23, F24, F27, F28, F41, F71, \& WAC-F71                                             & 10.4   &  756    &  14 \\  
NAC & 2014-08-01T18.31.34  & F82, F23, F24, F27, F28, F41, F71, \& WAC-F71                                             & 10.8   &  742    &  14 \\ 
NAC & 2014-08-01T19.37.34  & F82, F23, F24, F27, F28, F41, F71, \& WAC-F71                                             & 11.1   &  729    &  14 \\ 
NAC & 2014-08-01T20.43.34  & F82, F23, F24, F27, F28, F41, F71, \& WAC-F71                                             & 11.5   &  715    &  13 \\ 
NAC & 2014-08-03T00.21.16  & F82, F23, F24, F27, F28, F41, F71, \& WAC-F71                                             &  28.0  &  387    &  7.3 \\      
NAC & 2014-08-03T01.21.16  & F82, F23, F24, F27, F28, F41, F71, \& WAC-F71                                             &  29.1  &  377    &  7.1 \\     
NAC & 2014-08-03T02.21.16  & F82, F23, F24, F27, F28, F41, F71, \& WAC-F71                                             &  30.2  &  366    &  6.9 \\    
NAC & 2014-08-03T03.21.16  & F82, F23, F24, F27, F28, F41, F71, \& WAC-F71                                             &  31.6  &  356    &  6.7 \\     
NAC & 2014-08-03T04.21.16  & F82, F23, F24, F27, F28, F41, F71, \& WAC-F71                                             &  33.0  &  346    &  6.5 \\     
NAC & 2014-08-03T05.21.16  & F82, F23, F24, F27, F28, F41, F71, \& WAC-F71                                             &  34.4  &  336    &  6.3 \\    
NAC & 2014-08-03T06.21.16  & F82, F23, F24, F27, F28, F41, F71, \& WAC-F71                                             &  35.9  &  326    &  6.1 \\    
NAC & 2014-08-03T13.39.14  & F82, F23, F24, F27, F28, F41, F71, \& WAC-F71                                            &  40.7  &  287    &  5.4 \\
NAC & 2014-08-03T14.39.14  & F82, F23, F24, F27, F28, F41, F71, \& WAC-F71                                             &  40.7  &  284    &  5.3 \\     
NAC & 2014-08-03T15.39.14  & F82, F23, F24, F27, F28, F41, F71, \& WAC-F71                                             &  40.6  &  281    &  5.3 \\    
NAC & 2014-08-03T16.39.14  & F82, F23, F24, F27, F28, F41, F71, \& WAC-F71                                             &  40.6  &  278    &  5.2 \\
NAC & 2014-08-03T17.39.14  & F82, F23, F24, F27, F28, F41, F71, \& WAC-F71                                             &  40.6  &  275    &  5.2 \\
NAC & 2014-08-03T18.39.14  & F82, F23, F24, F27, F28, F41, F71, \& WAC-F71                                             &  40.6  &  273    &  5.1 \\ 
NAC & 2014-08-03T19.39.14  & F82, F23, F24, F27, F28, F41, F71, \& WAC-F71                                             &  40.6  &  270    &  5.1 \\
NAC & 2014-08-03T20.39.14  & F82, F23, F24, F27, F28, F41, F71, \& WAC-F71                                             &  40.6  &  267    &  5.0 \\
NAC & 2014-08-05T23.19.14  & F22, F23, F24, F27, F28, F41, F71, \& WAC-F71                                             &  48.9  &  123    &  2.3 \\   
NAC & 2014-08-06T00.19.14  & F22, F23, F24, F27, F28, F41, F71, \& WAC-F71                                             &  49.5  &  121    &  2.3 \\ 
NAC & 2014-08-06T01.19.14  & F22, F23, F24, F27, F28, F41, F71, \& WAC-F71                                             &   50.0 &  119    &  2.2 \\    
NAC & 2014-08-06T02.19.14  & F22, F23, F24, F27, F28, F41, F71, \& WAC-F71                                             &   50.8 &  117    &  2.2 \\
NAC & 2014-08-06T04.19.13  & F22, F23, F24, F27, F28, F41, F71, \& WAC-F71                                             &   52.2 &  113    &  2.1 \\     
NAC & 2014-08-06T05.19.14  & F22, F23, F24, F27, F28, F41, F71, \& WAC-F71                                             &   53.0 &  111    &  2.1 \\    
NAC & 2014-08-06T06.19.14  & F22, F23, F24, F27, F28, F41, F71, \& WAC-F71                                             &   53.9 &  110    &  2.1 \\ \hline \hline
\end{tabular}
}
\end{center}
\end{table*}


\section{Photometric properties from disk-averaged or integrated photometry}

\subsection{Phase function}

\begin{figure}[t]
   \centering
\includegraphics[width=9.2cm,angle=0]{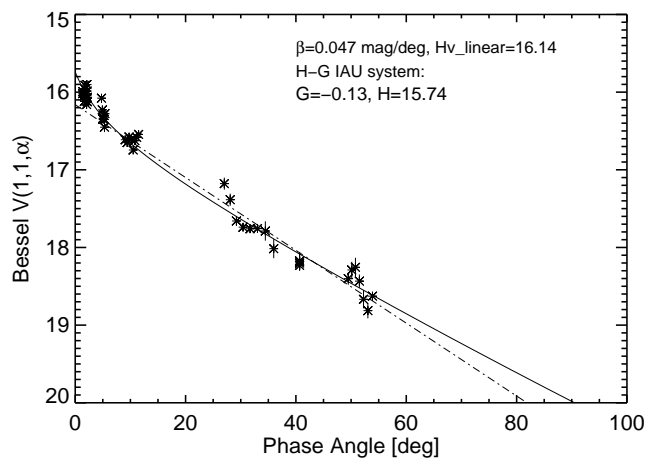}
      \caption{Phase curve of the nucleus of 67P from OSIRIS observations in the NAC green filter (F23). The continuous line represents the HG IAU model that best fits the data, and the linear fit to the data is plotted as the dashed-dotted
line.}
         \label{hg_func}
   \end{figure}

For an irregular body such as the nucleus of comet 67P, the definition
of the disk-integrated phase function becomes ambiguous (Li et al., 2004) because the observed disk-integrated reflectivity depends upon the solar phase
angle $\alpha$ and the illuminated surface seen during the observation. 
To built the phase function, we used the OSIRIS observations obtained with the NAC green filter, centered at 535 nm. Rosetta approached the comet very fast in July-August 2014, with a resulting rapid increase of the spatial resolution (see Table~\ref{observing}). The classical aperture photometry with a constant radius can
therefore not be applied to evaluate the comet flux. We decided to integrate the comet signal over the pixels with a flux higher than 0.2$\times$ $<F>$, where $<F>$ is the mean flux of the illuminated surface of the nucleus in a given image that was previously cut around the comet to avoid any potential background contributions (stars or cosmic rays). In this way, we integrated the comet flux over an area that is slightly larger than the projected surface of the comet obtained from the shape model for a given observing sequence, thus including  the limb contribution as
well.  We evaluated the coma contribution to be 0.1-0.2 \% of the comet signal in the green filter for the observations up to 3 August, and 0.2-0.5 \% for observations on 6 August, with the strongest coma contribution corresponding to the observation
made at 04:19, when a jet departing from the Hapi region (see Thomas et al., 2015, for the region nomenclature for comet 67P) is clearly visible in the stretched image (see Fig.~\ref{imareference}). The coma contribution is thus negligible given the low activity of the comet at these large heliocentric distances, and smaller than the uncertainties from absolute calibration, which are lower than 1\% for the green filter (Tubiana et al., 2015a). Nevertheless, we assumed a conservative approach for flux uncertainties, considering an error of 1.5\% of the flux to take into account the coma contribution and the absolute calibration uncertainties. 
We then corrected the measured comet flux to take into account the fact that the comet spectrum is redder than that of the Sun, using
\begin{equation}
F_c = F_o \times \frac{\int_\lambda F_{\odot}(\lambda)T(\lambda)\mathrm{d}\lambda}{\int_\lambda F_{comet}(\lambda)T(\lambda)\mathrm{d}\lambda}
,\end{equation}
where F$_c$ and F$_o$ are the corrected and uncorrected cometary fluxes at the central wavelength $\lambda_c$ of the green filter, T($\lambda$) is the system throughput (telescope optics and CCD quantum efficiency), and F$_{\odot}(\lambda)$ and $ F_{comet}(\lambda)$ are the solar (from the HST catalog) and cometary spectra (from Tubiana et al., 2011), respectively, both normalized to unity at the $\lambda_c$ of the green filter. \\
The visual magnitude of the comet was reduced to the standard Bessel V filter by
\begin{equation}
m_V = -2.5\log \left(F_c {\int_\lambda F_{comet} (\lambda)T_V(\lambda) \mathrm{d} \lambda}\right) + C
\label{mv}
,\end{equation}
where T$_V$ is the the transmission of the standard V filter taken from
Bessel (1990). We determined the C constant as the difference between
the apparent solar magnitude of –26.75 (Cox, 2000) and the convolution of the solar spectrum with the V filter as
in Eq.~\ref{mv}.
The absolute magnitude V(1,1,$\alpha$) reduced to the V-Bessel filter is shown in Fig.~\ref{hg_func}. 

We used the HG system (Bowell et al., 1989) to characterize the phase-function behavior of the nucleus and its absolute magnitude,
H$_v$(1,1,0). Our best-fit values for both parameters are G=-0.13$\pm$0.01, and $H_v(1,1,0)$ = 15.74$\pm$0.02 mag, which were determined from chi-squared fitting.
The nucleus displays a strong opposition effect:
its linear slope (calculated for $\alpha > 7^o$) is $\beta$ = 0.047$\pm$0.002 mag/$^{\circ}$, and the magnitude from the linear slope (thereby excluding the opposition effect) is H$_{Vlin}$ = 16.16$\pm$0.06 mag. The scatter of the measurements obtained at similar phase angles (Fig.~\ref{hg_func}) is related to the rotational phase of the comet.

Attempts to determine the phase function of the nucleus of  67P
were made using ground-based observations as well. 
Tubiana et al. (2011) sampled the phase-angle range 0.5$^{\circ}$--10$^{\circ}$ and concluded that a linear approximation better represents their measurements than the IAU-adopted phase function. They determined a very steep magnitude dependence on phase angle, with linear phase coefficients in the range 0.061-0.076 mag/$^{\circ}$.
Lamy et al. (2007) combined HST (Lamy et al., 2006) and ESO-NTT observations (Lowry et al., 2006) and obtained a slope parameter G=-0.45, which implies a very steep brightness dependence on phase angle and a possible strong opposition effect. This result might be affected by changes in the observing geometry that are
due to the orbital motion of the comet.
Lowry et al. (2012) determined a slope parameter G=0.11 $\pm$ 0.12 and, by a linear fitting, $\beta$ = 0.059 $\pm$ 0.006 mag/$^{\circ}$. \\
The phase function determined from the OSIRIS observations has a lower linear slope than the values previously obtained from ground-based and HST observations, simply because it is evaluated for phase angles $> 7^o$, excluding the contribution from the
opposition effect. If we consider a linear fit of the OSIRIS data in the 1-10$^{\circ}$ range, then the linear slope is 0.074 mag/$^{\circ}$, thus compatible with previous measurements. Ciarniello et al. (2015) used data from the VIRTIS instrument and found a steeper slope for $\alpha < 15^o$ (0.082$\pm$0.016 mag/$^{\circ}$) and a lower value (0.028$\pm$0.001 mag/$^{\circ}$) than  we found for $25 < \alpha < 110^{\circ}$. The discrepancies of the results between OSIRIS and VIRTIS data are related to the different phase-angle range used to compute the slope and to the fact that Ciarniello et al. (2015) derived the integrated magnitude from the disk-averaged albedo assuming that the nucleus  of 67P is spherical with an equivalent radius of 1.72 km, while we directly measured the integrated flux. \\
The linear slope we determined is close to the average value for Jupiter-family comets ($\beta$ = 0.053$\pm$0.016 mag/$^{\circ}$, Snodgrass et al., 2011), very similar to the value found for comets Hartley 2 ($\beta$=0.046 mag/$^{\circ}$, Li et al., 2013), Tempel 1 ($\beta$=0.046 mag/$^{\circ}$, Li et al., 2007a, 2013), and Borelly ($\beta$=0.043 mag/$^{\circ}$, Li et al., 2007b), and similar to that of low-albedo asteroids (Belskaya \& Shevchenko 1999).

\subsection{Disk-averaged spectrophotometry} 
 
   \begin{figure}[t]
   \centering
 \includegraphics[width=9.2cm,angle=0]{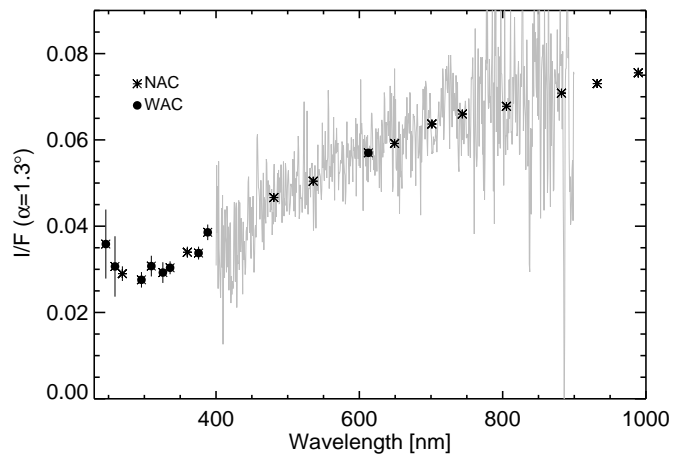}
      \caption{Average spectrophotometry of NAC and WAC observations obtained on 28 July, 23:25-23:45 UT. The continuous gray line represents a ground-based spectrum of  67P  taken in 2007, when the comet was inactive (Tubiana et al., 2011).}
         \label{spec_all}
   \end{figure}
   
 67P  has been observed with several NAC filters since the very first resolved images, allowing us to investigate the spectrophotometric properties of the nucleus. We present in this section the global spectrophotometric properties derived from the NAC and WAC images obtained on 28 and 29 July, 2014, when Rosetta was at a distance of about 2000 km from the comet, at the smallest phase-angle reached during the 2014 observations ($\alpha$=1.3$^{\circ}$). The corresponding spatial resolution was 37 m/px for the NAC  (Table~\ref{observing}). 
The data set includes sequences that consist of observing blocks of 11 NAC filters and 2 WAC filters repeated nine times during the rotational period of the comet (from 28 July 14:39 UT to  29 July 00:45 UT ), but, as a result of pointing uncertainties, only the last three sequences, starting from 22:19 UT, are centered on the comet and useful for our analysis. In addition, and only for this run, we obtained parallel observing sequences including all the 14 WAC filters repeated eight times from 28 July 15:35 UT to 29 July 00:55 UT. Unfortunately, the WAC images acquired with filters covering $\lambda>$ 500 nm are saturated or have fluxes beyond the linearity range of the detector,  therefore we decided to discard them from our analysis.
We derived a mean I/F value for each NAC and WAC image by integrating the cometary signal for all the pixels with a flux $> 0.5 \times$ $<I/F>$, where $<I/F>$ is the mean radiance factor of the illuminated pixels on the cometary\ surface, and finally dividing by the projected surface of the comet derived from the stereophotogrammetric (SPG) shape model (Preusker et al., 2015). \\
The spectrophotometry  derived from the  WAC and NAC images taken on 28 July, 23:25 UT, is shown in Fig.~\ref{spec_all}, with a superposed ground-based spectrum of  67P taken from Tubiana et al. (2011). The OSIRIS spectrophotometry in the visible and near-infrared range is fully consistent with ground-based observations, as also summarized in Table~\ref{ground_slope}. \\
 In general, the nucleus has similar spectrophotometric properties in the NUV-VIS-NIR range to those of bare cometary nuclei (Lamy et al., 2004), of primitive D-type asteroids such as Jupiter Trojans (Fornasier et al., 2007, 2004), and of the moderately red Transneptunians and Centaurs (Fornasier et al., 2009).  No clear absorption bands are visible at the spectral resolution of the filters we used, and the spectral slope may change beyond 750 nm, being fainter in the NIR range than in the VIS. However, this change of slope might not be real because the 700 nm and 743 nm fluxes are  contaminated by the coma emissions present at this large heliocentric distance (3.65 AU), as we show in Sect. 4.3. 
 
Figure~\ref{spec_all} shows that the flux rises up in the mid-UV (240-270 nm range) for the two WAC filters F31 and F41 and for the NAC-F15. These two WAC filters, designed to study the CS emission and the adjacent continuum at 245 nm, are very difficult to calibrate because they are affected by pinholes (Tubiana et al., 2015a), therefore some additional light may contaminate the signal. However, we checked the background signal for the two filters, and it is not as high as expected for a significant flux contribution from pinholes defects. If we assume that this signal excess is at least partially real, then a tentative explanation of this broad band, which has a minimum close to 290 nm, may be SO$_2$ frost. A similar feature was seen on Jupiter satellites,
for instance, the trailing hemisphere of Europa, where it was attributed to sulphur dioxide ice (Noll et al., 1995).\ Another
similar feature was observed on comet Bowell 1980b (A'Hearn et al., 1984) and on the Centaur 2060 Chiron; Brosch (1995) attributed
it to water ice or other frost such as NH$_3$. \\
Laboratory experiments on cryogenic films of SO$_2$ and SO$_2$/H$_2$O show that the SO$_2$ absorption band, usually located close to 280 nm,  is irreversibly temperature dependent (Hodyss et al., 2013) and it moves toward longer wavelength with the increase of the temperature. Hodyss et al. (2013) found a minimum close to 287 nm for T $\sim$ 140 K, and this minimum must slightly shift to longer wavelengths for higher temperatures, such as those measured on  67P  by the VIRTIS instrument, that is, 180-230 K (Capaccioni et al., 2015).
For T $>$ 160 K, the $\rm SO_2$ can sublimate and thus the gas sulphur dioxide will be photolyzed into SO+O and 
$\rm S + O_2$ with relatively high photodissociation coefficients (in the range of $1 \times 10^{-4}$ and $5 \times 10^{-5}$ 
$\rm s^{-1}$ at 1 AU). It must be noted that the ROSINA instrument has identified sulphur compounds in the coma of 67P (Altwegg et al., 2015), and that the 67P UV spectra acquired with ALICE are compatible with the presence of SO$_2$ ice (Feaga et al., 2015). Therefore, the absorption feature at $\sim 290$ nm could be due to SO$_2$ or any other ice containing that molecule in its structure.

We also see an excess of the flux in the OH filter centered at 309 nm compared to the nearby continua (Fig.~\ref{spec_all}).
To understand whether this higher flux at 309 nm is due to the resonance fluorescence of OH produced by water photolysis,
we have, on one hand, studied the radial profiles of the emission both at 309 nm and in the nearby continua, and on the other
hand, studied the remaining OH emission after subtracting the underlying continuum measured from the observation in the  UV295 filter. \\
From the study of the radial profiles measured from the observations at 295 nm and 309 nm, no conclusive results can be drawn because 
both profiles show a similar behavior, which prevents concluding that the behavior  of OH is characteristic of a gas species 
produced by a parent species. We also produced pure OH gas maps by subtracting the underlying continuum contribution.
The resulting OH image is such that the emission excess is on the same order as the background signal. 
However, although the analysis is not conclusive because the
images obtained on 28-29 July were not designed to study gas emissions, 
it is worth noting that H$_2$O in the gas phase was detected since early June 2014 by the MIRO instrument (Gulkis et al., 2015), showing periodic variations related to nucleus rotation and shape.  OSIRIS already detected an outburst of activity in April 2014 (Tubiana et al., 2015b), and the ROSINA instrument (Altwegg et al., 2015) also measured H$_2$O. This implies that
the OH radical, coming from water photodissociation, is probably also present in the coma. \\
Figure~\ref{spec_all} also shows a slight increase of the I/F value at wavelengths where CN fluorescent emission takes place (i.e., 387 nm).  A similar analysis to the one made for the OH case indicates that, if any CN surrounds the nucleus, the emission is on the same order of magnitude as that of the background. Thus, a clear confirmation of CN presence in the coma of 67P at these heliocentric distances cannot be drawn from our data.

\begin{table*}
\begin{center}
\caption{Spectral slope of the nucleus of 67P, according to the
spectral slope definition of Delsanti et al. (2001), derived from the literature and the data presented here.}
\label{ground_slope}
\begin{tabular}{l l c c c } 
Date & Slope (\%/100nm) & Wavel. range (nm) & $\alpha$ [$^{\circ}$] & Reference \\ \hline \hline
April 2004 & 12 $\pm$ 2 & 545 - 797 & 2.1 & Tubiana et al., 2008 \\
June 2004 & 5 $\pm$ 3 &  545 - 797 & 10.3 & Tubiana et al., 2008 \\
May 2006 & 11 $\pm$ 2 &  545 - 797 & 0.5-1.3 & Tubiana et al., 2008 \\
May 2006 & 11 $\pm$ 2 & 500 - 850 & 0.5-1.3 & Tubiana et al., 2008 \\
July 2007 &  10 $\pm$ 2 & 545 - 797 & 6.0 & Tubiana et al., 2011 \\
July 2007 & 11 $\pm$ 1 & 500 - 850 & 6.0 & Tubiana et al., 2011 \\
July 2007 &  10 $\pm$ 1 & 550 - 650 & 6.0 & Lowry et al., 2012 \\
21 July 2014  & 11.2$\pm$0.3     & 535-882   & 4.8-5.4 & This work \\  
25 July 2014  & 11.3$\pm$0.3     & 535-882   & 2.2 & This work \\
28-29 July 2014  &  11.5$\pm$0.5 & 535-882   & 1.3 & This work \\   
1 Aug. 2014  &  12.2$\pm$0.3 & 535-882   & 9.0-11.5 & This work \\   
3 Aug. 2014  &  14.7$\pm$0.7 & 535-882   & 28-36 & This work \\   
3 Aug. 2014  &  15.6$\pm$0.9 & 535-882   & 40-41 & This work \\   
6 Aug. 2014  &  15.8$\pm$0.4 & 535-882   & 50-54 & This work \\  \hline
\\
\end{tabular}
\end{center}
\end{table*}

\subsection{Phase reddening}

\begin{figure}[h]
   \centering
\resizebox{\hsize}{!}{\includegraphics{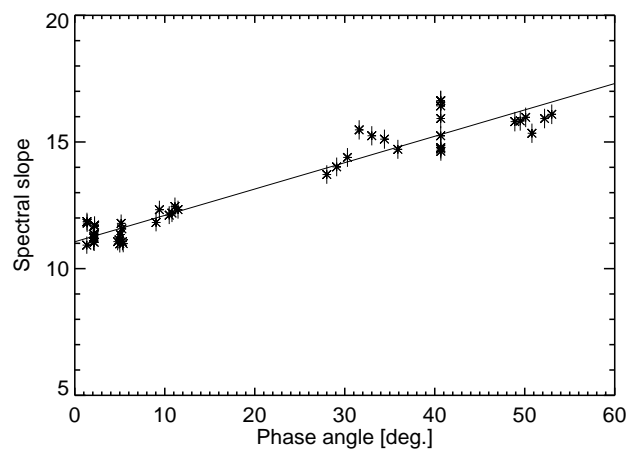}}      
\caption{Variation of the mean spectral slope of the nucleus
of 67P, evaluated  in the 882-535 nm range and expressed in \%/(100 nm), with the phase angle.}
         \label{reddening}
   \end{figure}

We computed the disk-averaged spectral slope ($S$) following the definition of Delsanti et al. (2001) as \\
\[ S = \frac{R_{882} - R_{535}}{R_{535} \times (882 ~nm - 535 ~nm),} \]
where R$_{882}$, and R$_{535}$ are the mean comet I/F values in the filters centered on 882 nm and 535 nm. 
The mean spectral slope evaluated in the 882-535 nm range is shown in Fig~\ref{reddening}.  The obtained spectral slope varies between 11\%/(100 nm) at a phase angle of 1.3$^{\circ}$ and 16\%/(100 nm) at a phase angle of 52$^{\circ}$, implying a significant phase reddening. This is the first time that significant phase-reddening effects are detected on a comet nucleus. Strong phase-reddening effects on comet 67P are also seen from the VIRTIS data both in the visible and near-infrared range (Ciarniello et al., 2015). Li et al. (2007a) used a different spectral slope definition, that is, it is normalized in the center of the wavelength range, and found a red slope of 12.5$\pm$1.0 \% for comet Tempel 1 from the Deep Impact mission observations at a phase angle
of 63$^{\circ}$ and no sign of phase reddening in the two high-phase angles they sampled (63$^{\circ}$ and 117$^{\circ}$). With the same spectral slope definition, Li et al. (2013) measured a spectral slope of 7.6$\pm$3.6\%/(100 nm), and of 8.6$\pm$10.6\%/(100 nm) for comet Hartley 2 at phase angles of 85$^{\circ}$ and 93$^{\circ}$, respectively. As a result of the large uncertainties in the spectral slope values, no conclusions on the phase reddening of Hartley 2 can be drawn.\\
If we compute the spectral slope applying the same method as Li et al. (2007, 2013), that is, normalizing the reflectance value at the central wavelength range instead that at 535 nm, we find a slope value of 12.5\%/(100 nm) at 52$^{\circ}$ that is similar to that of comet Tempel 1.

The linear fit that best matches our data has a slope $\gamma = $ 0.104$\pm$0.003 $\times$ 10$^{-4}$ nm$^{-1}/^{o}$, and the disk-averaged spectral slope is estimated to be 11.3$\pm$0.2 \%/(100 nm) at zero phase angle. \\ The phase reddening was first noted by Gehrels et al. (1964) on the lunar
surface, and it is potentially attributed to the increased contribution
of the multiple scattering at large phase angles as the wavelength and
albedo increase, plus a contribution of surface roughness effects (Sanchez et al., 2012; Hapke et al., 2012; Schroder et al., 2014).
As reported in  Fig.~\ref{reddening}, the mean spectral slope varies not only with the phase angle, but, at similar phase angle, also with the rotational phase. Higher spectral slope values correspond to redder surfaces, that is, when bluer regions like Hapi are not visible. For example, the peak in the spectral slope at $\alpha$=32-33$^{\circ}$ corresponds to color sequences taken on 3 August at UT 03:21-04:22 (Fig.~\ref{color1}) when the Hapi region, which is the bluest region of the comet (see Sect. 4.2), was not visible at all.

\subsection{Hapke model on disk-averaged photometry}

\begin{center}
\begin{table*}
\caption{\label{tab:Hapke} Hapke parameters derived from disk-averaged reflectance for different 
filters.}
\begin{centering}
{\small{}}%
\begin{tabular}{cccccccc}
\hline 
 {\small{$\lambda$(nm)}} & {\small{$w_{\lambda}$}} & {\small{$g_{\lambda}$}} & {\small{$B_{0}$}} & {\small{$h_{s}$}} & {\small{$\bar{\theta}$}} & {\small{$A_{geo}$}} & {\small{$A_{bond}$}} \tabularnewline
\hline 
\hline 
  {\small{325}} & {\small{$0.028\pm0.001$}} & {\small{$-0.35\pm0.03$}} & {\small{$1.83\pm0.03$}} & {\small{$0.029\pm0.005$}} & {\small{$15$}} & {\small{$0.0316\pm0.0030$}} & {\small{$0.0088\pm0.0007$}}  \tabularnewline
  {\small{480}} & {\small{$0.035\pm0.001$}} & {\small{$-0.43\pm0.03$}} & {\small{$1.91\pm0.09$}} & {\small{$0.021\pm0.004$}} & {\small{$15$}} & {\small{$0.0554\pm0.0024$}} & {\small{$0.0119\pm0.0001$}} \tabularnewline
 {\small{535}} & {\small{$0.037\pm0.002$}} & {\small{$-0.42\pm0.03$}} & {\small{$1.95\pm0.12$}} & {\small{$0.023\pm0.004$}} & {\small{$15$}} & {\small{$0.0589\pm0.0034$}} & {\small{$0.0123\pm0.0001$}}  \tabularnewline
  {\small{649}} & {\small{$0.045\pm0.001$}} & {\small{$-0.41\pm0.03$}} & {\small{$1.97\pm0.09$}} & {\small{$0.026\pm0.007$}} & {\small{$15$}} & {\small{$0.0677\pm0.0039$}} & {\small{$0.0157\pm0.0001$}} \tabularnewline
  {\small{700}} & {\small{$0.050\pm0.001$}} & {\small{$-0.38\pm0.02$}} & {\small{$2.22\pm0.06$}} & {\small{$0.027\pm0.002$}} & {\small{$15$}} 
& {\small{$0.0720\pm0.0031$}} & {\small{$0.0173\pm0.0001$}} \tabularnewline
  {\small{743}} & {\small{$0.053\pm0.004$}} & {\small{$-0.40\pm0.03$}} & {\small{$1.96\pm0.07$}} & {\small{$0.023\pm0.003$}} & {\small{$15$}} & {\small{$0.0766\pm0.0031$}} & {\small{$0.0178\pm0.0001$}} \tabularnewline
  {\small{882}} & {\small{$0.052\pm0.003$}} & {\small{$-0.40\pm0.04$}} & {\small{$2.08\pm0.09$}} & {\small{$0.027\pm0.006$}} & {\small{$15$}} & {\small{$0.0780\pm0.0038$}} & {\small{$0.0179\pm0.0001$}} \tabularnewline
{\small{989}} & {\small{$0.066\pm0.005$}} & {\small{$-0.35\pm0.03$}} & {\small{$2.07\pm0.05$}} & {\small{$0.032\pm0.004$}} & {\small{$15$}} & {\small{$0.0820\pm0.0039$}} & {\small{$0.0223\pm0.0001$}} \tabularnewline
\hline 
\end{tabular}
\par\end{centering}{\small \par}
\end{table*}
\par\end{center}
 
 
We computed the disk-averaged reflectance in different filters by integrating the cometary signal for all the pixels with a flux $> 0.5 \times$ $<I/F>$, where $<I/F>$ is the mean radiance factor of the illuminated pixels on the cometary surface, and finally dividing by the projected surface of the comet derived from the SPG shape model. \\
The disk-averaged multiwavelength phase curve (in radiance factor, $I/F$)
in the 1.3$^{\circ}$--54$^{\circ}$ phase-angle range allows us to study the photometric properties of the comet's surface. We used the Hapke model (Hapke, 1993),
a semi-theoretical assembly of functions that describes the contribution
of single-scattering, multiple-scattering, opposition effect and roughness
to reproduce the measured reflectance. The model has experienced several improvements with more physical description since its introduction
in the 1980s (Hapke, 1993, 2002, 2008, 2012), which is the reason
that there are several different versions. For a comet that is expected to be composed of dark
phases rich in carbon, the coherent-backscattering mechanism (Shevchenko
\& Belskaya, 2010; Shevchenko et al., 2012) is not expected to play a major role. 
For the single-particle-phase function, we adopted the single-term of the Henyey-Greenstein (HG) form (Hapke 2012), with one asymmetry parameter, $g$. A negative $g$ value represents backscattering, a positive one forward scattering, and $g=0$ stands for isotropic scattering. 

Since the first resolved images, the nucleus of 67P presented a highly irregular shape. However, Li et al. (2003)
have reported that the influence of shape is irrelevant when obtaining the Hapke parameters when the data are fit with phase angles lower than 60$^{\circ}$,
which is our regime. Therefore, we adopted a five-parameter
version of the model, built for spherical bodies, that includes the single-scattering albedo $w_{\lambda}$,
the asymmetry factor $g_{\lambda}$, the average roughness angle $\bar{\theta}$,
and the amplitude $B_{0}$ and width $h_{s}$ of the shadow-hiding
opposition surge. The disk-integrated Hapke expression for the radiance
factor can be written as
\begin{equation}
\footnotesize{
\begin{split}
\frac{I_{\lambda}}{F_{\lambda}}=K(\alpha,\bar{\theta})\biggl[\left(\frac{w_{\lambda}}{8}
\left[(1+B_{SH_{\lambda}}(\alpha))P_{hg}(\alpha,g_{\lambda})-1\right]
+\frac{r_{0\lambda}}{2}\left(1-r_{0\lambda}\right)\right)\\ 
\left(1-\sin\frac{\alpha}{2}\tan\frac{\alpha}{2}\ln\left[\cot\frac{\alpha}{4}\right]\right)
+\frac{2}{3\pi}r_{0\lambda}^{2}(\sin\alpha+(\pi-\alpha)\cos\alpha)\biggl],
\end{split}
}
\end{equation}

where $\alpha$ is the phase angle, $\lambda$ is the wavelength were the $\frac{I}{F}$ is measured, and $r_0$ is defined as

\begin{equation}
r_{0\lambda}=\frac{1-\sqrt{1-w_{\lambda}}}{1+\sqrt{1-w_{\lambda}}},
\end{equation}
where $w$ is the single scattering albedo. The shadow-hiding opposition effect $B_{SH_{\lambda}}(\alpha)$ and the shadowing function $K_{\lambda}(\alpha,\bar{\theta})$ are
given by Hapke (1993):

\begin{equation}
B_{SH_{\lambda}}(\alpha,B_{0},h_{s})=\frac{B_{0}}{1+\frac{\tan\frac{\alpha}{2}}{h_{s}}},
\end{equation}

\begin{equation}
K_{\lambda}(\alpha,\bar{\theta})=\exp\left\{ -0.32\bar{\theta}\left[\tan\bar{\theta}\tan\frac{\alpha}{2}\right]^{1/2}-0.52\bar{\theta}\tan\bar{\theta}\tan\frac{\alpha}{2}\right\} 
,\end{equation}
\noindent
and the single-term Henyey-Greenstein function is expressed as
\begin{equation}
P_{hg}(\alpha,g)=\frac{(1-g_{\lambda}^{2})}{(1+2g_{\lambda}\cos\alpha+g_{\lambda}^{2})^{3/2}}
.\end{equation}

To model the observed data with the Hapke formula, we developed a procedure that consists of randomly picking
30 initial conditions for each parameter for a given filter, and we searched for the global minimum between the observations and the fit using Broyden-Fletcher-Goldfarb-Shanno (BFGS) algorithm (Byrd et al., 1995) in the Basin-Hopping method (Wales \& Doye, 1997). The minimum was searched for inside the boundaries $w_{\lambda}=\left\{ 0.02,0.1\right\} ,\,\, g_{\lambda}=\left\{ -1.0,1.0\right\} ,\,\, B_{0}=\left\{ 0.2,2.5\right\} , \text{and}\,\, h_{s}=\left\{ 0.0,0.5\right\} $. 
Preliminary tests showed that $\bar{\theta}$  is ill constrained due to the lack of points at phase angles larger than 60$^{\circ}$, so we decided to fix it to $15^{\circ}$, the value determined from the model of Hapke (2002) on the disk-resolved images, as discussed in Sect. 4.2.

The Basin-Hopping method has been used to solve complex molecular systems (e.g., Prentiss et al., 2008; Kim et al., 2014)
and is based on scanning several local minima to estimate the global minimum. When the solution was found, we
ran a Levenberg-Marquardt algorithm of the MINPACK library (More,
1978) for each parameter individually to adjust them to a better precision. 
The best solution was chosen from the 30 tests as the test that returned the
lowest $\chi^{2}$ value. The standard deviation of the parameters was
calculated from the spread of the solutions of each test.
Because we used a method that estimates the global solution, 
deviations among the  solutions found give an estimate of the error bars of the different parameters. Undersampling and correlations among the Hapke parameters (Bowell et al., 1989) are the main sources of errors when flux uncertainties are negligible. \\
The obtained phase curve shows a steeper behavior starting from $\sim\alpha=$5$^{\circ}$, probably because of the shadow-hiding
opposition effect. The obtained amplitude of the opposition
effect is no smaller than 1.78, and its width is no lower than 0.023. The comet has thus a sharper opposition effect than that found from an average phase function for C-type asteroids ($B_{0}=1.03$ and $h_{s}=0.025$, Helfenstein \& Veverka, 1989).

\begin{figure}[t]
   \centering
\includegraphics[width=8.5cm]{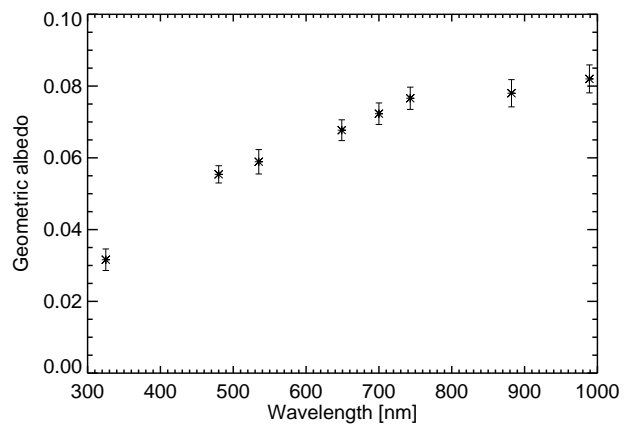}
      \caption{Geometric albedo for different filters derived from  Hapke modeling of the disk-averaged photometry.}
         \label{geom_albedo}
   \end{figure}

The Hapke parameters we obtained for the 67P disk-average reflectance (Table
\ref{tab:Hapke}) are very similar to those found by Li et al. (2012) on disk-averaged reflectance at 650 nm of comet Tempel 1 from the Deep Impact data 
(i.e., $w_{650nm} = 0.043\pm0.006$, $g_{\lambda}=-0.48\pm0.02$, $\bar{\theta}=20\pm5$ and $A_{geo}=0.059\pm0.046$ ). 
We report all multifilter Hapke solutions from the disk-averaged reflectance in Table~\ref{tab:Hapke}. 
Hapke modeling also enables deriving the geometric albedo, which
is reported in Fig.~\ref{geom_albedo} for the different wavelengths
we investigated.
The Hapke parameters show no clear wavelength dependence of $g_{\lambda}$,
$B_{0}$ and $h_{s}$, within the uncertainties. This behavior is expected when the shadow-hiding
effect is the main cause of the opposition surge, while for the
asymmetry factor, it means that the average scattering path of light
inside the surface particles is small and that light is rapidly
absorbed.

\section{Disk-resolved photometry, colors, and spectrophotometry}

           \begin{figure*}
   \centering
 \includegraphics[width=15cm,angle=0]{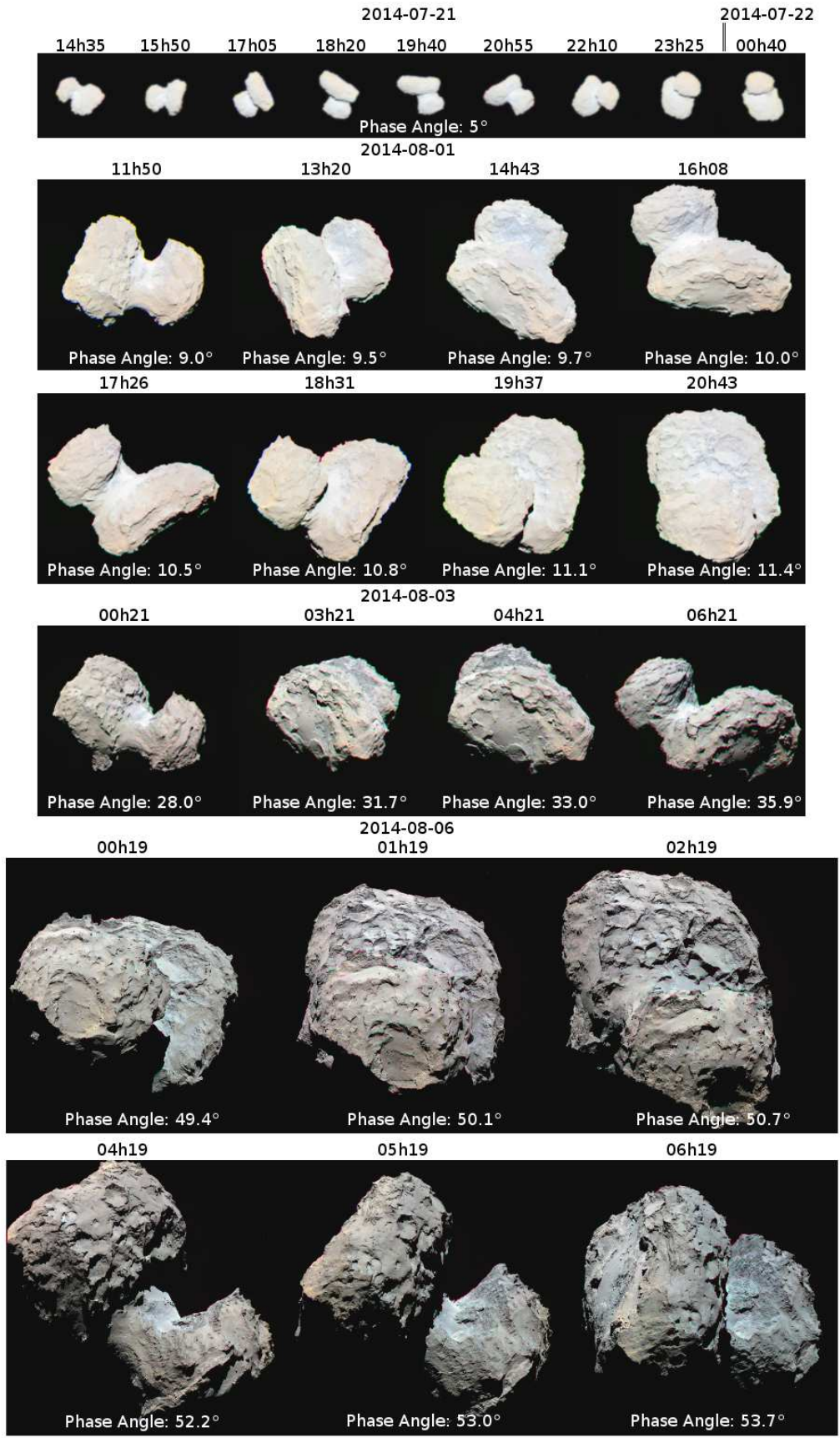}
      \caption{RGB images, in false colors, of the nucleus of 67P from different observing runs. The color images are produced using the filters centered on 480 nm, 649 nm, and 882 nm. The spatial resolution varies from 90 m/px on July 21 (at the top) to 2.1 m/px on 6 August (at the bottom).}
         \label{color1}
   \end{figure*}

         \begin{figure*}
   \centering
 \includegraphics[width=18cm,angle=0]{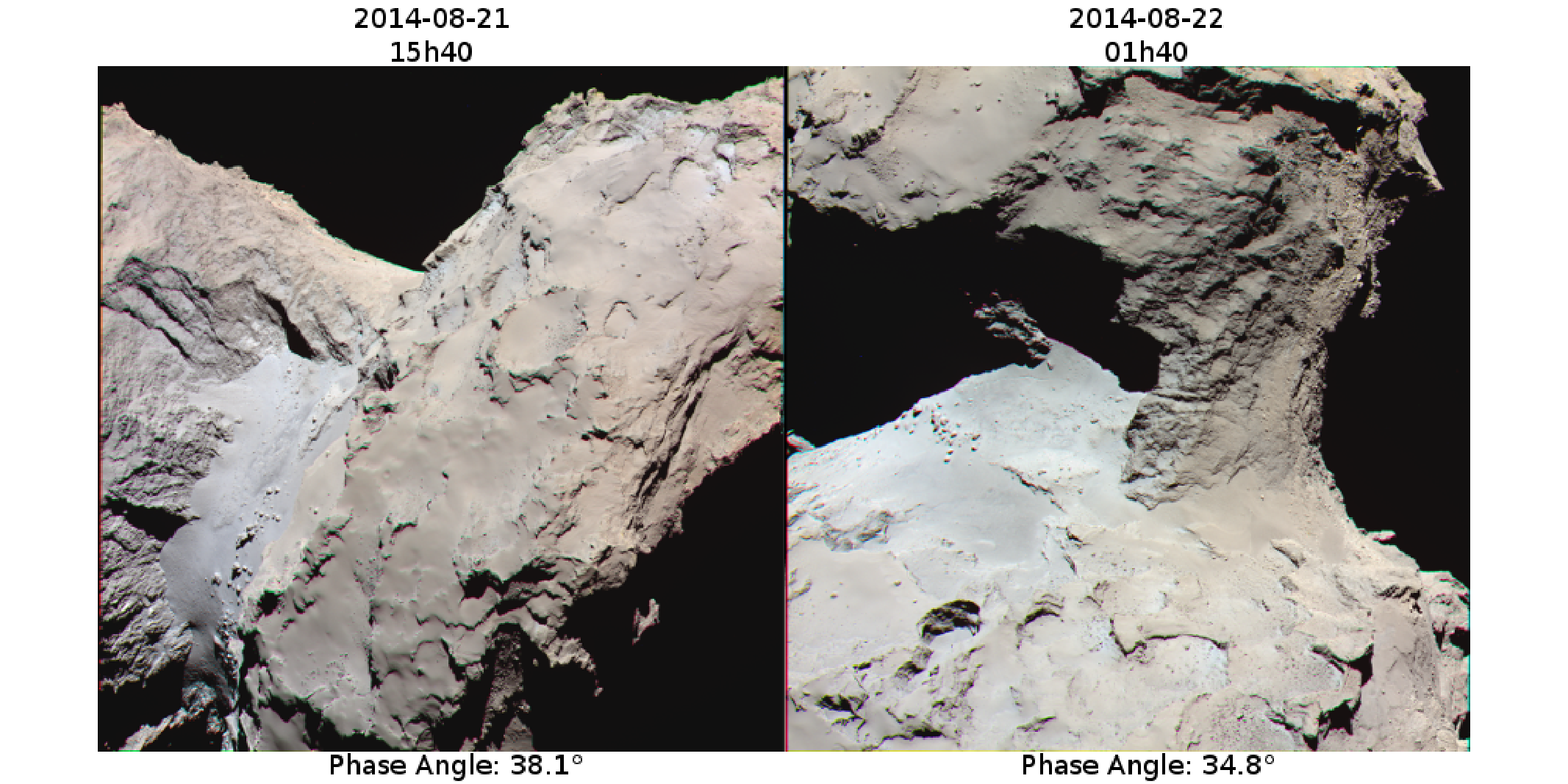}
      \caption{RGB images, in false colors, of the nucleus of 67P from images taken on 21 and 22 August with the filters centered on 480 nm, 649 nm, and 882 nm. Rosetta was at a distance of 70 km from the nucleus, and the corresponding resolution is 1.3 m/px.}
         \label{color2}
   \end{figure*}

Along with the OSIRIS NAC images, we derived additional files that contain geometric information about the illumination and observation angles. For each NAC image pixel, we computed solar incidence, emission, and phase-angle information. This computation considers all relevant geometric parameters, such as the camera distortion model, the alignment of the camera to the Rosetta spacecraft, the orientation of the spacecraft (with reconstructed orbit position and pointing) with respect to the 67P nucleus, and a 3D shape model of the comet. The shape model and the improved spacecraft orientation data are based upon a stereophotogrammetric analysis (SPG) of OSIRIS NAC images. SPG has already been successfully applied to OSIRIS NAC images taken during the Rosetta flyby of the asteroid (21) Lutetia (Preusker et al., 2012) and to imaging observations of many other planetary missions. The SPG-based shape model of the nucleus of 67P  used in this analysis consists of about 6 million vertices (about 12 million facets) and comprises a sampling distance of about 2 m on the  nucleus surface (Preusker et al., 2015). \\
Color cubes of the surface were produced by stacking registered and illumination-corrected ($(I/F)_{corr}$) images. For the illumination correction, we used a simple Lommel-Seeliger disk law,
\begin{equation}
D(i,e) =  \frac{2 \cos(i)}{\cos(i) + \cos(e)}
.\end{equation} 
The Lommel-Seeliger law comes from radiative transfer theory when considering a single-scattering particulate surface (Fairbairn, 2005). This law has been proved to be suitable for smooth low-albedo surfaces, such as C-type asteroids and cometary nuclei, due to the predominance of single scattering, especially at low phase angle, as predicted by radiative transfer models (Hapke, 1981). Disk-corrected images are hence obtained as
\begin{equation}
(I/F)_{corr}(\alpha, \lambda) = \frac{\pi I(i,e,\alpha,\lambda)/(F_{\lambda})}{D(i,e)}
.\end{equation}

\subsection{Color variations}

OSIRIS data show the 67P surface with unprecedented resolution compared to other cometary nuclei visited by space missions.  Figures~\ref{color1} and ~\ref{color2} show the RGB maps, in false colors, generated using NAC images acquired from 25 July to 22 August with the filters centered on 882 nm, 649 nm, and 480 nm. Images were first co-registered, then RGB images were generated using the STIFF code (Bertin, 2012), a software developed to produce color images from astronomic FITS data. 
 
The nucleus of comet 67P has a complex shape, with different types of terrains and morphological features (Thomas et al., 2015), and it shows some color variations, in particular in the Hapi region (Figs.~\ref{color1} and ~\ref{color2}). This region is located between the two lobes of the comet, it is both the brightest and the most active surface on the comet at the large heliocentric distances monitored during the approach phase and bound orbits of Rosetta. Hapi is also the source of the most spectacular jets seen at heliocentric distances $>$ 3 AU (Lara et al., 2015).  Other bright regions appear at the boundary between the Aten and the Ash area together with bright, circular features on the Seth unit. At higher resolution, several bright spots associated with exposed water ice are seen in all types of geomorphologic regions on the comet, and these features are fully discussed in Pommerol et al. (2015). \\
The peculiar colors of some regions of the comet, especially for the Hapi region, are confirmed by the bluer spectral slope compared to the darker regions, as we discuss in Sect. 4.2. Moreover, the Hapi region itself shows difference in colors when observed at higher resolution (Fig.~\ref{color2}), and these differences are observable through different filters and within multiple phase angles and resolutions, hence they are independent of observational geometries and shape corrections.

 \subsection{Albedo map and disk-resolved photometry}

         \begin{figure}[h]
   \centering
 \includegraphics[width=8.7cm,angle=0]{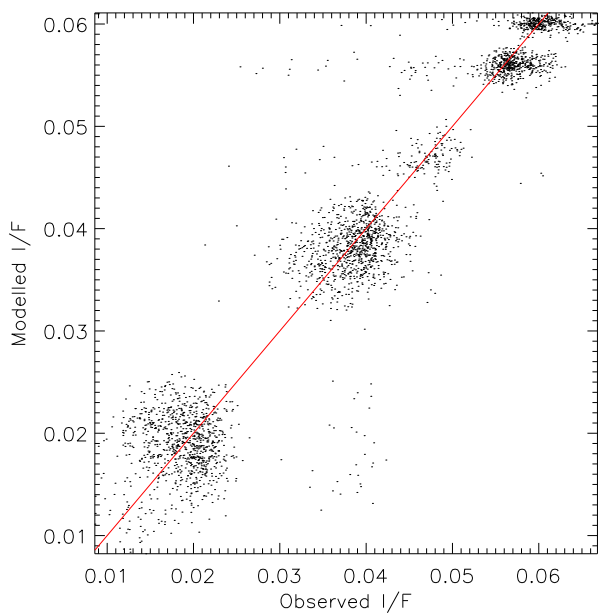}
      \caption{Goodness plot of the modeled $I/F$ with respect to measured $I/F$ (at 649 nm) of Hapke (2002) modeling of the nucleus of 67P.}
         \label{hapkefit}
   \end{figure} 
   \begin{figure}[t]
   \centering
 \includegraphics[width=9.2cm,angle=0]{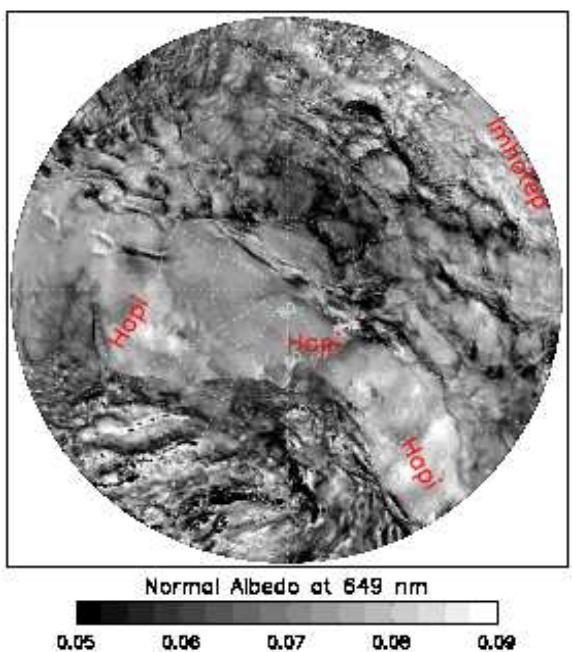}
      \caption{Normal albedo polar map of the nucleus of 67P in the orange filter (649 nm). The longitude 0$^o$ in the Cheops reference frame is to the bottom (Preusker et al., 2015).}
         \label{albedomap}
   \end{figure}   
We investigated the local photometry and reflectance variations of the observed surface of the nucleus at 649 nm using Hapke's (2002) reflectance  model. The disk-resolved data offer many combinations of emission, incidence, and azimuth angles, allowing us to constrain some parameters in a better way than using disk-integrated data. 
We used topographic information based on the SPG shape model, together with the observed reflectance at 649 nm on all the illuminated pixels of the selected images described in Table~\ref{observing}.
In the Hapke (2002) model, the bidirectional reflectance ($Ref$) at a given incidence (i), emission (e), and phase angle ($\rm \alpha$) is described as 

\begin{equation}
{\small
Ref(i,e,\alpha) = \frac{\omega}{4\pi} \frac{\mu_0}{\mu_0 + \mu} [P_{hg}(\alpha,g)B_{SH}(g)+M(\mu_0,\mu)]B_{CB}(g),
}
\end{equation}
where $\mu_0~=~\cos(i)$, $\mu~=~\cos(e)$, $P_{HG}$ is the single-term Henyey-Greenstein function,  $B_{SH}$ the shadow-hiding opposition term, $\rm B_{CB}$ the coherent-backscattering term, and M the multiple-scattering term described in Hapke (2002, Eq. 17).
To provide equivalent weight to the observations obtained at different phase angles, all the images were first binned at an interval of about 2.0 degrees into a 3D grid composed of the three geometric angles. The I/F pixels were averaged over each bin. In this way, we removed the influence of intrinsic albedo variations on the fitting. To avoid unfavorable observation geometries due to large incidence and emission angles, we filtered the data and considered pixels with incidence and emission angles lower than 70$^o$. This constrain excludes extreme geometries near the limb or near the terminator. A least- $\chi^2$ fit using the Levenberg-Marquardt algorithm was performed to find the global Hapke parameters, giving an RMS error of 6\%. The modeled $I/F$ with respect to the measured $I/F$ values are shown in Fig.~\ref{hapkefit}.
The Hapke parameters that best fit our data are given in Table~\ref{hapkedisk2}. 

\begin{table*}
\caption{Results from the Hapke (2002, and 2012) modeling from disk-resolved images taken with the NAC orange filter centered on 649 nm.}
\label{hapkedisk2}
\begin{centering}
\begin{tabular}[t]{ccrrrrrrrrr}
\hline \\
Model & $w$  & $g$ & $B_{0SH} $ & $h_{s}$  & $\bar{\theta} ~[^{\circ}$] & $B_{0CB}$ & $h_{CB}$ & Porosity & K & Geom. alb. \\
\hline \\
Hapke 2002 &  0.042 & -0.37 & 2.5 & 0.079 & 15 & 0.188 & 0.017 & - & - & 0.064 \\
\hline\\
Hapke 2012 & 0.034 & -0.42 & 2.25 & 0.061 & 28  & -&- & 0.87 & 1.2 & 0.067 \\
\hline \\
\end{tabular}
\end{centering}
\end{table*}

To test the robustness of the results, we also computed the Hapke parameters using the latest version of the Hapke model (Hapke, 2012), which also takes into account the top layer porosity, given by the porosity factor $K= - \ln(1-1.209 \phi^{2/3})/(1.209 \phi^{2/3})$, where $\phi$ is the filling factor, and the porosity is defined as 1-$\phi$. The complete equation for this version of the model can be found in Helfenstein \& Shepard (2011).
We fit all images together using the Broyden-Fletcher-Goldfarb-Shanno (BFGS) algorithm (Byrd et al., 1995), which solves nonlinear problems by approximating the first and second derivatives in an iterative procedure to search for a local minimum.  \\
The boundaries used for the parameters are $w_{\lambda}=\left\{ 0.02,0.1\right\} ,\,\, g_{\lambda}=\left\{ -1.0,1.0\right\} ,\,\, B_{0}=\left\{ 0.2,2.5\right\} ,\,\, h_{s}=\left\{ 0.0,0.5\right\} $, and $\bar{\theta}=\left\{ 5^{o},90^{o}\right\} $.
We decided against using the basin-hopping method in this case because of constraints in the convergence time. Nevertheless, we used the disk-integrated results as initial conditions to the BFGS algorithm, expecting that the best solution might be situated around the same minimum. \\
At the end of the procedure, we observed that the coherent-backscattering terms have remained invariant to the final solution,  so we can rule out the coherent-backscattering process as a significant effect to the opposition regime at least for $\alpha > 1.3^{\circ}$. 
The best solutions from Hapke (2012) modeling are reported in Table~\ref{hapkedisk2}. We observe that the regolith's particles are backscattering, with a $g$ parameter value close to the one found from the disk-integrated analysis. The single-scattering albedo value is lower than the one found from the disk-integrated analysis or from the Hapke (2002) model. This is due to the inclusion of the porosity term (Hapke 2008, 2012) that also acts on the opposition effect and multiple-scattering formulae. To compare the single-scattering albedo from the two models, the $w_{\lambda}$ from the Hapke (2012) solution must therefore be multiplied by the porosity factor K, which has a value of 1.2 for a porosity of 0.87. In doing so, we obtained a Hapke (2012) porosity-corrected $w$ value of 0.041, very close to the $w$ value found from the Hapke (2002) model. \\
We find a porosity value reaching 87\% for the 67P nucleus top surface layer, which is higher than the 40-60\% value found by the KOSI experiments (Sears et al., 1999; Kochan et al., 1998) that simulated cometary top surfaces. Laboratory experiments also showed the formation of a thick dust mantle on cometary-analog materials that may be exposed to dust settling and pulverization, facilitating the formation of a very weak and porous mantle (Mohlmann, 1995; Sears et al., 1999). Moreover, recent works have evoked fractal aggregates (Bertini et al., 2007; Levasseur-Regourd et al., 2007; Lasue et al., 2010) as the best analogs of cometary dust. Surfaces composed of such material have a porosity ranging from 80\% to 90\%, in agreement with what we found from the Hapke model.

The single-scattering albedo, roughness angle, and geometric albedo found from Hapke (2002, 2012) modeling are very similar to the values found for comets 9P/Tempel 1 and 81P/Wild 2  (Li et al., 2009, 2012). The asymmetry factor found for comet 67P is lower than the one derived for comets Tempel 1, Hartley 2, and Wild 2 (Li et al., 2007a, 2009, 2013), implying that the regolith on comet 67P is less backscattering than for these comets. Ciarniello et al. (2015) analyzed disk-resolved data of comet 67P from VIRTIS observations and found an asymmetry factor of -0.42, a roughness of 19$^{\circ}$, and a single-scattering albedo of 0.052$\pm$0.013 at 550 nm. While the  $\theta$ and $g$ values they determined are similar to the one we found, the single-scattering albedo is much higher than the values we determined both from the disk-averaged and disk-resolved analysis of the OSIRIS data at 535-649 nm. It must be noted that Ciarniello et al. (2015) used a simplified version of the Hapke model, neglecting the opposition-effect terms and the multiple-scattering functions (implying an overestimation of the single-scattering albedo), which they applied to VIRTIS data covering the 27-111$^{\circ}$ phase-angle range. The different formalism and wavelength range covered explain the discrepancy in the single-scattering albedo values.\\ 
If we compare the Hapke parameters from disk-resolved and disk-integrated photometry (Tables~\ref{tab:Hapke} and ~\ref{hapkedisk2}), we see that the integrated solution gives a similar $w$ but much lower $h_s$ values than the disk-resolved solutions. This discrepancy is related to the different formalisms applied to compute the parameters, and to the fact that the effects of the different parameters are entangled in the disk-integrated function. Nonetheless, without data under 1.3 degrees of phase angle, it is difficult to precisely determine the amplitude of the opposition effect. 
Therefore, Hapke parameters obtained from the disk-resolved data should be preferred as they are computed by modeling the full brightness variations across the comet surfaces. 
It is noteworthy that despite the discrepancies between disk-integrated and disk-resolved results, we obtained very similar geometric albedo values at 649 nm from the different solutions (Tables~\ref{tab:Hapke} and ~\ref{hapkedisk2}).
       \begin{figure*}[t]
   \centering
 \includegraphics[width=16.0cm,angle=270]{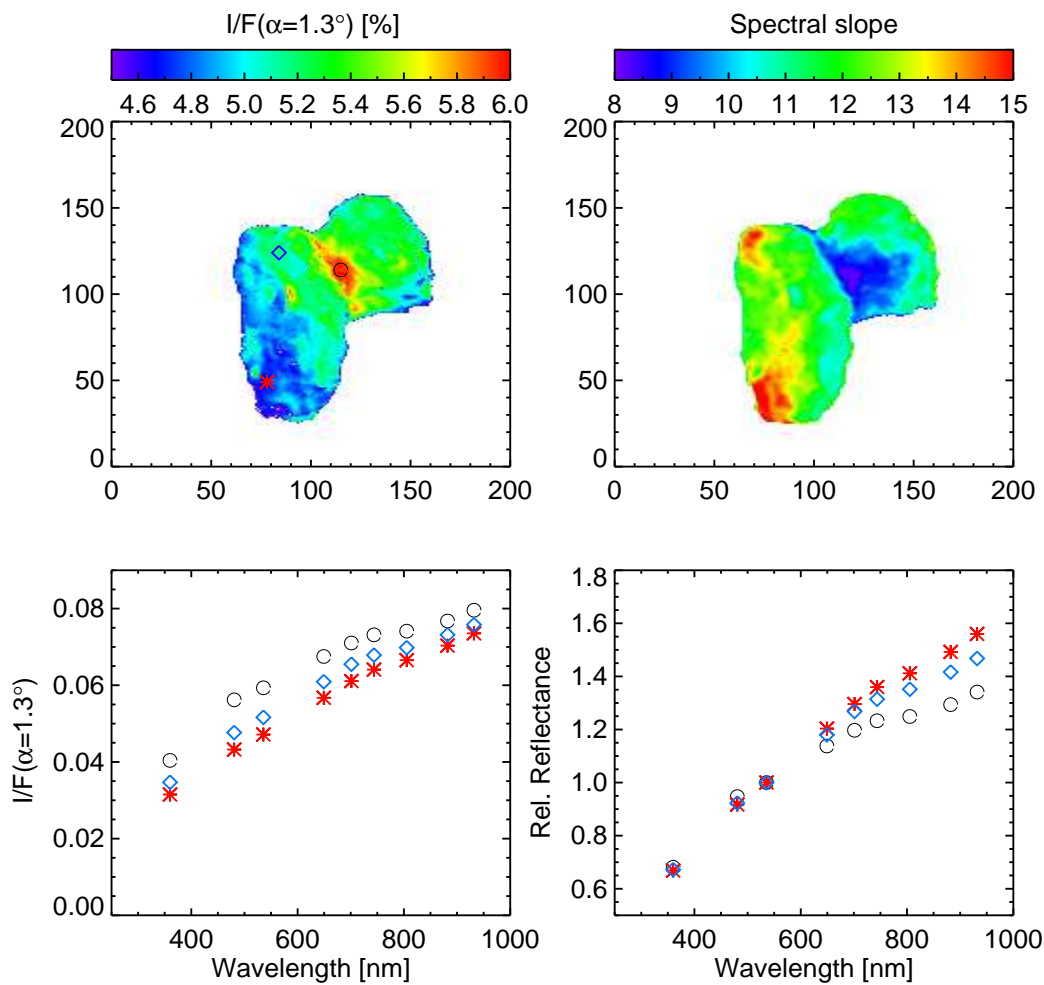}
      \caption{I/F for the green filter centered on 535 nm, slope map, absolute and relative reflectance for some selected regions on the nucleus of 67P for the 29 July observations taken at 00:43 UT and at $\alpha = 1.3^{\circ}$. The slope is computed in the 535-882 nm range, after normalization at 535 nm, and it is in \%/(100 nm). The I/F map and the reflectance are corrected for the illumination conditions using the Lommel-Seeliger disk law.}
         \label{slopemap}
   \end{figure*}
%


The albedo map shown in Fig.~\ref{albedomap} was then generated using a rotational movie sequence obtained on 3 August in which most of the comet surface was observed several times at very different incidences and emission angles. We used the parameters derived from the Hapke (2002) model to correct the reflectance in the following manner:\\

\begin{equation}
I/F_{\rm corrected}=I/F_{measured}\frac{I/F_{Hapke}(i_{ref},e_{ref},\alpha_{ref}) } {I/F_{Hapke}(i_{obs},e_{obs},\alpha_{obs})}
,\end{equation}

where $I/F_{\rm measured}$ stands for the observed reflectance at the given incidence, emission, and phase angles
$(i_{obs},e_{obs},\alpha_{obs})$, while $i_{ref}, e_{ref}$, and $\alpha_{ref}$ are the geometric angles at which the reflectance $I/F_{corrected}$ is supposed to be. Here we have chosen $i_{ref}$=0$^{\circ}$, $e_{ref}$=0$^{\circ}$, and $\alpha_{ref}$=0$^{\circ}$, so the corrected I/F should be the normal albedo. \\
The $I/F_{\rm corrected}$ of each surface element (facet) was then computed and stored as soon as a facet was illuminated. For each facet we obtained multiple measurements that were finally averaged to compute the normal albedo. Approximately 50 images were used to generate the albedo map represented in Fig.~\ref{albedomap}. The same approach was used by Leyrat et al. (2010) to study the photometric properties of asteroid Steins from the OSIRIS data acquired during the Rosetta flyby.

             \begin{figure*}[t]
   \centering
 \includegraphics[width=18.5cm,angle=0]{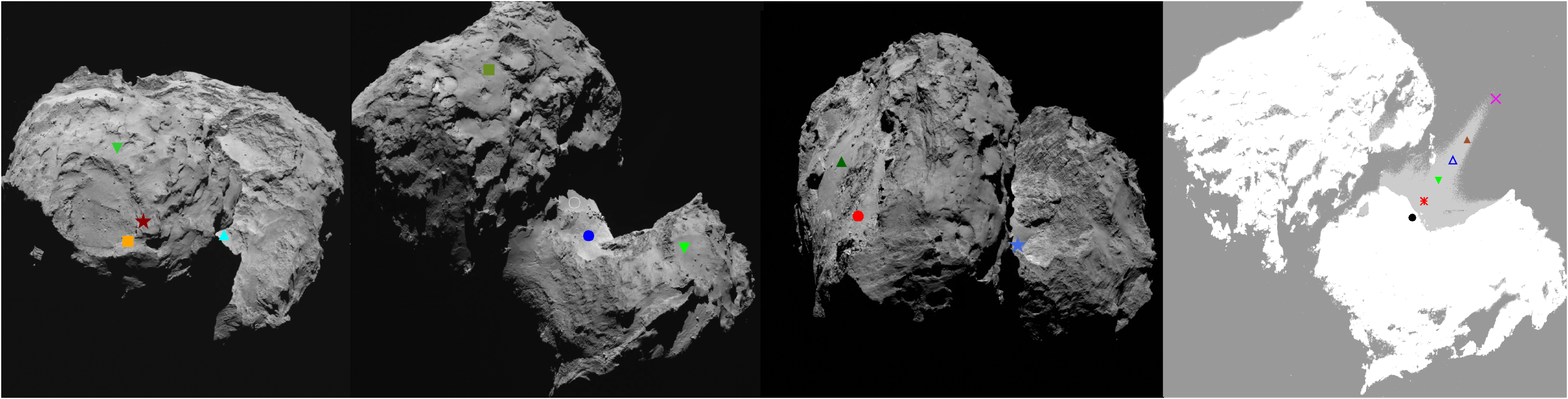}
      \caption{Three images acquired on 6 August at 00:19, 04:19, 06:19 UT, and again 04:19 UT, from left to right. The first three images from the left side show the selected regions where the spectrophotometry on the nucleus has been computed, as represented in Fig.~\ref{Relativereflectance}. The last image on the right side is taken at 04:19 UT but stretched to show the coma features. The points in this image represent different nucleus and coma regions whose flux is represented in Fig.~\ref{xx}.}
         \label{imareference}
   \end{figure*} 

The comet nucleus is dark and the estimated geometric albedo from the Hapke parameters is 6.5$\pm$0.2\% at 649 nm. This value is slightly higher but still similar to that found for comets Wild 2 and Tempel 1 (see Table 1 in Li et al., 2013 for a summary of the photometric parameters of comet nuclei visited by space missions). 
The histogram of the albedo distribution derived from the albedo map shown in Fig.~\ref{albedomap} was fit with a Gaussian function that peaks at 0.063 and has a standard deviation $\sigma$ of 0.01. Thus the surface reaches albedo variations of $\sim~$30\% within $2 \sigma$ of the albedo distribution.\\ Some very bright spots in Fig. 8 seem to be related to high-frequency topographic features that are only poorly described by the shape model (i.e., surrounding terrains of the Imhotep region and cliffs between Hapi and Hathor). Overall, Hapi is the brightest region, also
showing some local albedo variations, as already noted in the RGB images (Fig.~\ref{color2}). The Imhotep area also shows a significant albedo increase over a wide surface. Both the Hapi and Imhotep regions have a smooth surface, suggesting that the bright albedo may be at least partially due to the texture of the surface. On the other side, the Seth geological area is the darkest of the northern hemisphere.

\subsection{Local spectrophotometry and composition} 

 We computed the spectral slope for each pixel in the 535-882 nm range on co-registered and illumination corrected images as \\
\[ S = \frac{Rc_{882}(\alpha) - Rc_{535}(\alpha)}{Rc_{535}(\alpha) \times (882 ~nm - 535 ~nm),} \]
where Rc$_{882}(\alpha)$, and Rc$_{535}(\alpha)$ are the comet I/F values in a given pixel, corrected for the illumination conditions using the Lommel-Seeliger disk law, in the filters centered on 882 nm and 535 nm. 
Images were normalized to the green filter, centered on 535 nm, to be consistent with most of the literature data on primitive solar system bodies whose spectral slopes are usually computed on spectra normalized in the V filter or at 550 nm. We decided to correct the images for the illumination conditions using the simple Lommel-Seeliger disk law and not using the Hapke formalism simply because the Hapke parameters from disk-resolved images are only available for the filter centered on 649 nm, and not at several wavelengths. \\
 Figure~\ref{slopemap} shows the I/F map, illumination corrected using the Lommel-Seeliger disk law, of the comet obtained on 29 July with the NAC green filter, at the lowest phase angle (1.3$^{\circ}$) sampled during the 2014 observations, as well as the spectral slope map, evaluated in the 882-535 nm range, and the relative and absolute reflectance of three regions selected over the surface. The disk-corrected image at 535 nm has a mean I/F value of 5.2\% at $\alpha=1.3^{\circ}$, and shows albedo variations up to +16\% in the brightest Hapi region, and -9\% in the darker Ash region. The reflectance is anticorrelated with the spectral slope, and in particular, the bright Hapi region is also characterized by a bluer spectral slope than the darker
medium-reflectance regions.  We computed the absolute and relative spectrophotometry in a box 3 px $\times$ 3 px wide (i.e., $110~m \times 110~m$) on three selected regions representing dark, medium, and high reflectance areas (red star, blue squares, and black circles  in Fig.~\ref{slopemap}). The selected dark area shows a linear red
slope without obvious spectral features, and it has a steeper spectral slope. The bright area selected in the Hapi region has a different spectral behavior, bluer than the other two regions, with a clear change of slope beyond 650 nm and possibly an absorption feature in the 800-900 nm range. All the three regions show an enhancement of the fluxes at 700--743 nm due to cometary emissions that we describe in the following paragraphs. \\
      \begin{figure}
   \centering
 \includegraphics[width=9.0cm,angle=0]{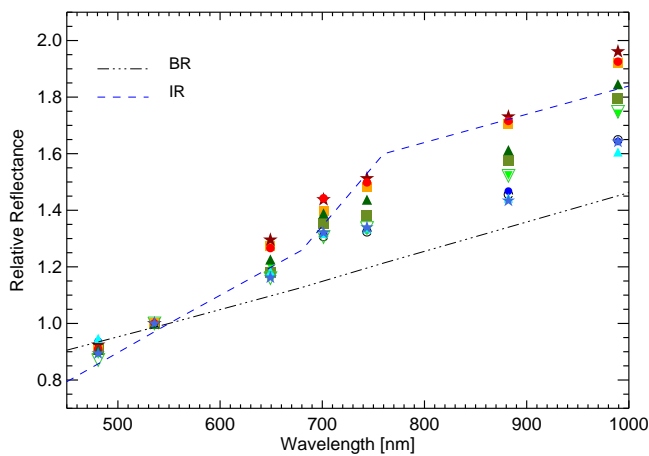}
      \caption{Relative reflectance from ten selected regions as displayd in Fig.~\ref{imareference}. The dashed-dotted and dashed lines represent the mean spectrophotometry for the TNO taxonomic classes BR and IR (Barucci et al., 2005), respectively.}
         \label{Relativereflectance}
   \end{figure}
 \begin{figure}
   \centering
 \includegraphics[width=9.0cm,angle=0]{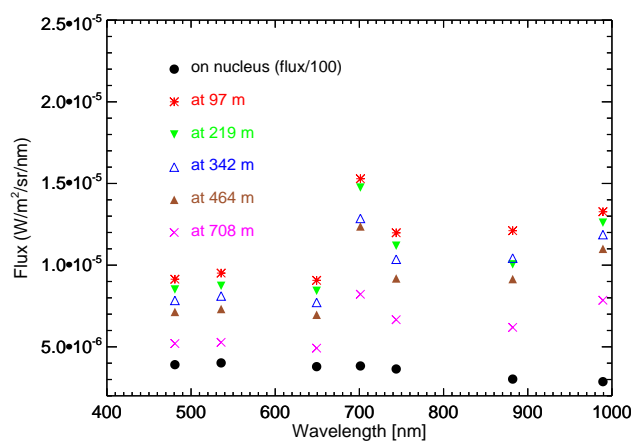}
      \caption{ Flux of the comet nucleus (divided by 100) and of the coma at several projected distances (the selected regions are represented in Fig.~\ref{imareference}, last image on the right) from the nucleus surface for the observations on 6 August acquired at 04:19 UT.}
         \label{xx}
   \end{figure}
We then studied the spectrophotometry of the comet obtained at higher spatial resolution during the observations on 6 August. 
Figure~\ref{Relativereflectance} shows the relative reflectance, normalized at 535 nm and averaged over a box 5$\times$5 pixels wide ($\sim 10 \times 10~m^2$) for ten selected regions on the nucleus as reported in Fig.~\ref{imareference}.  To study the relative spectrophotometry, we sampled three different groups of terrains, from the brightest and spectrally less red Hapi region to the reddest surfaces on the comet (Figs.~\ref{imareference} and ~\ref{Relativereflectance}). The brightest regions, in particular Hapi, are clearly less red in term of spectral slope. We interpret this spectral behavior as being caused by a higher abundance of water ice at the surface, on the basis of the correlation between bluer slope and ice abundance found for  comets Hartley 2 and Tempel 1 (Li et al., 2013; Sunshine et al., 2006, 2013), and on the basis of the results from experiments on ice or dust mixture sublimation (Pommerol et al., 2015). Moreover, the Hapi region is often shadowed by the Hathor region, it is not illuminated at perihelion due to seasonal effects, and thus it is a favored region to retain ices on its surface. It must be noted that no clear water ice absorption bands have been detected with the VIRTIS infrared-imaging spectrometer during the August 2014 observations at resolutions of 15-25 m/px (Capaccioni et al., 2015), even though the shape of the strong absorption beyond 3 $\mu$m in the Hapi region seems to be consistent with the presence of low amounts of water ice. The absence of large areas of water-ice-rich mixtures in VIRTIS data is attributed to the lower spatial resolution of these data than that of the OSIRIS images, together with the presence of dark-phased and non-volatile materials that may mask the water ice spectral absorptions. 
\\
The reddest  spectral regions are found in both lobes of the comet, in particular nearby the Hatmehit depression in the small lobe and in the Apis region in the large lobe. The different spectral behaviors of the selected regions over the nucleus are comparable with those of other cometary nuclei, with Jupiter Trojan asteroids and with part of the transneptunian (TNO) objects (Fornasier et al., 2007, 2009; Lamy et al., 2004) . We also plot in Fig.~\ref{Relativereflectance} the two TNO taxonomic classes BR and IR (defined in Barucci et al., 2005) whose spectrophotometry is close to that observed in different areas of the comet. These two classes are intermediate between very blue and very red TNOs.

With the spectral resolution given by the  filters used it is not clear whether some absorptions features are present in the near-infrared regions, and the interpretation is complicated by the fact that we clearly see an excess of flux in the 700-750 nm region. To ascertain the origin of this flux excess, we studied the behavior of this emission at several locations both on the nucleus and in the coma. More concisely, for the images obtained
on 6 August, we selected several locations on the nucleus, immediately (as projected on the sky) above the surface, at the inner limb, and in the coma to average the radiance in squares of $ 10 \times 10$ m$^2$, as was done for the analysis of different regions over the nucleus, for every filter. These measurements were made both in cometary regions where jet-like features are clearly visible and where they seem to be absent. \\ 
Figure \ref{xx} shows the coma radiance on the nucleus and at different projected distances, from oservations on 6 August acquired at UT 04:19.  Whereas at $\lambda < 649.2$ nm, the I values on the nucleus and in the the dust coma follow the same trend, at 701 nm there is a noticeable increase in the inner coma locations, and the peak values decrease with increasing projected distance from the nucleus surface. A similar behavior is observed at 743 nm and tentatively at $\lambda \ge 880$nm, although the latter cannot be clearly confirmed as there are no data acquired with adjacent filters to allow any emission band isolation.  On the directly illuminated nucleus surface both the band-like structure at 701-743 nm and the I/F strong
increase in the near-IR cannot be detected when compared with the nucleus spectrum itself. That this behavior is mainly
observed at coma positions indicates that this flux is unrelated to the nucleus surface composition, but is related to the coma. The enhancement of the flux in the 700-750 nm region must then be associated with the cometary emissions in the coma.  Furthermore, it was also verified that it can be clearly seen in shadowed areas, where the strong nucleus  signal is absent. \\
Using Mie-scattering models, we have ruled out that dust grains were responsible for the  high values of I at 701 nm. Thus, the other possible candidates are emissions produced by gas species in the coma of 67P. 
From high-resolution spectra of comets (Cochran \& Cochran, 2002), we conclude that the most likely candidates to give rise to such an enhanced emission are $\rm H_2O^+$ in its (0,6,0) band at 700 nm, the (0,2,0)-(0,0,0) band at 743-745 nm, and the band system (0,6,0) of $\rm NH_2$ that overlaps with the water bands.  Water, the parent species of  $\rm H_2O^+$, was reported by Hassig et al. (2015), and Gulkis et al. (2015) from data acquired by the  ROSINA and MIRO instruments, respectively, as early as August 4, 2014. The processes by which $\rm H_2O$ is ionized once is still under study, although electronic impact and photoionization are among the most plausible ones.  \\
In addition to this coma feature, the spectrally bluer  to moderately red regions show a rise of the flux at 989 nm (Fig.~\ref{Relativereflectance}), and this behavior was also seen in the  VIRTIS spectra (Capaccioni et al., 2015). In particular, the spectrally bluer regions like Hapi seem to present a faint absorption feature centered between 
800 nm and 900 nm, which may be associated to the Fe$^{2+}$ -- Fe$^{3+}$ charge transfer absorption band in silicates. Unfortunately, the low spectral resolution of the OSIRIS observations and the contamination of the adjacent filters in the 700-750 nm regions by coma emissions complicates the interpretation of this potential absorption feature. 

   \begin{figure*}
   \centering
 \includegraphics[width=11cm,angle=-90]{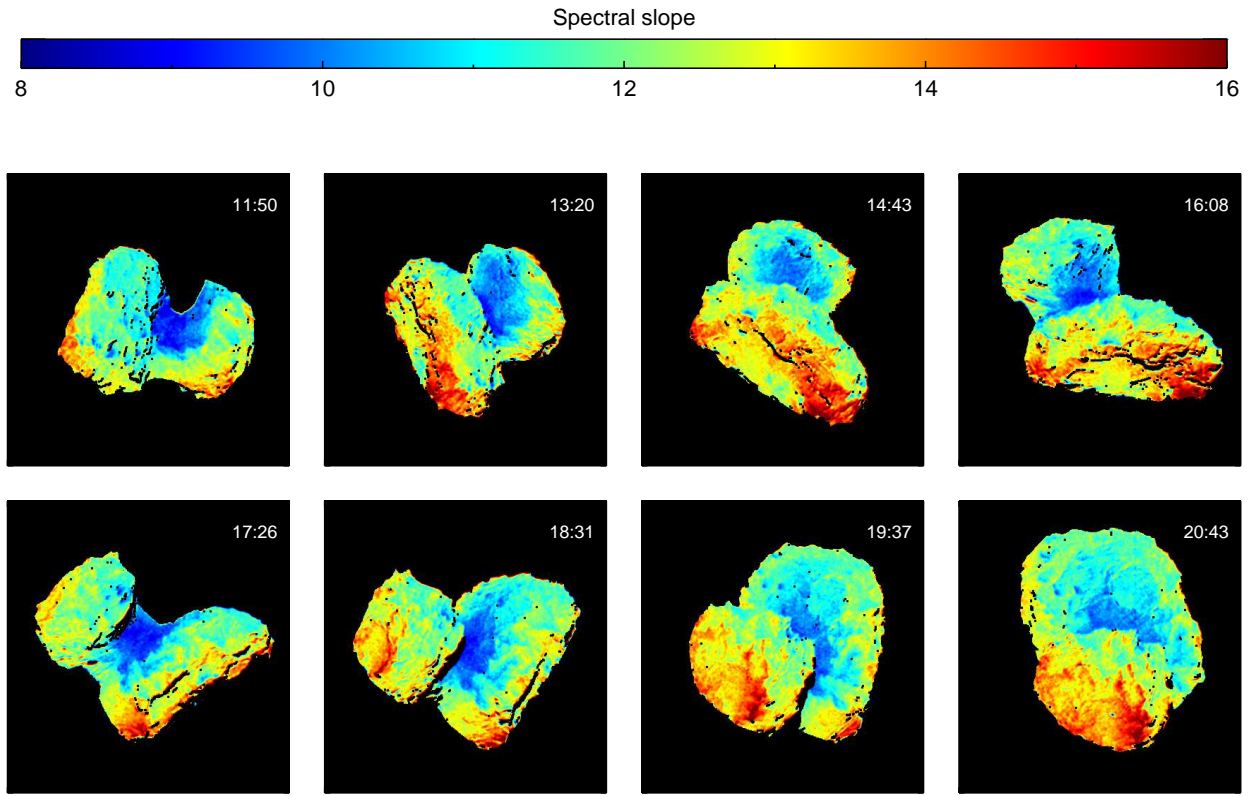}
      \caption{Spectral slope images for the observations on 1 August, at a phase angle of 9-10$^{\circ}$. The slope is computed in the 535-882 nm range, after normalization at 535 nm, and it is in \%/(100 nm).}
         \label{specmap_1aug}
   \end{figure*}

\subsection{Correlations between albedo, color variations, and geology}

      \begin{figure*}
   \centering
 \includegraphics[width=11cm,angle=-90]{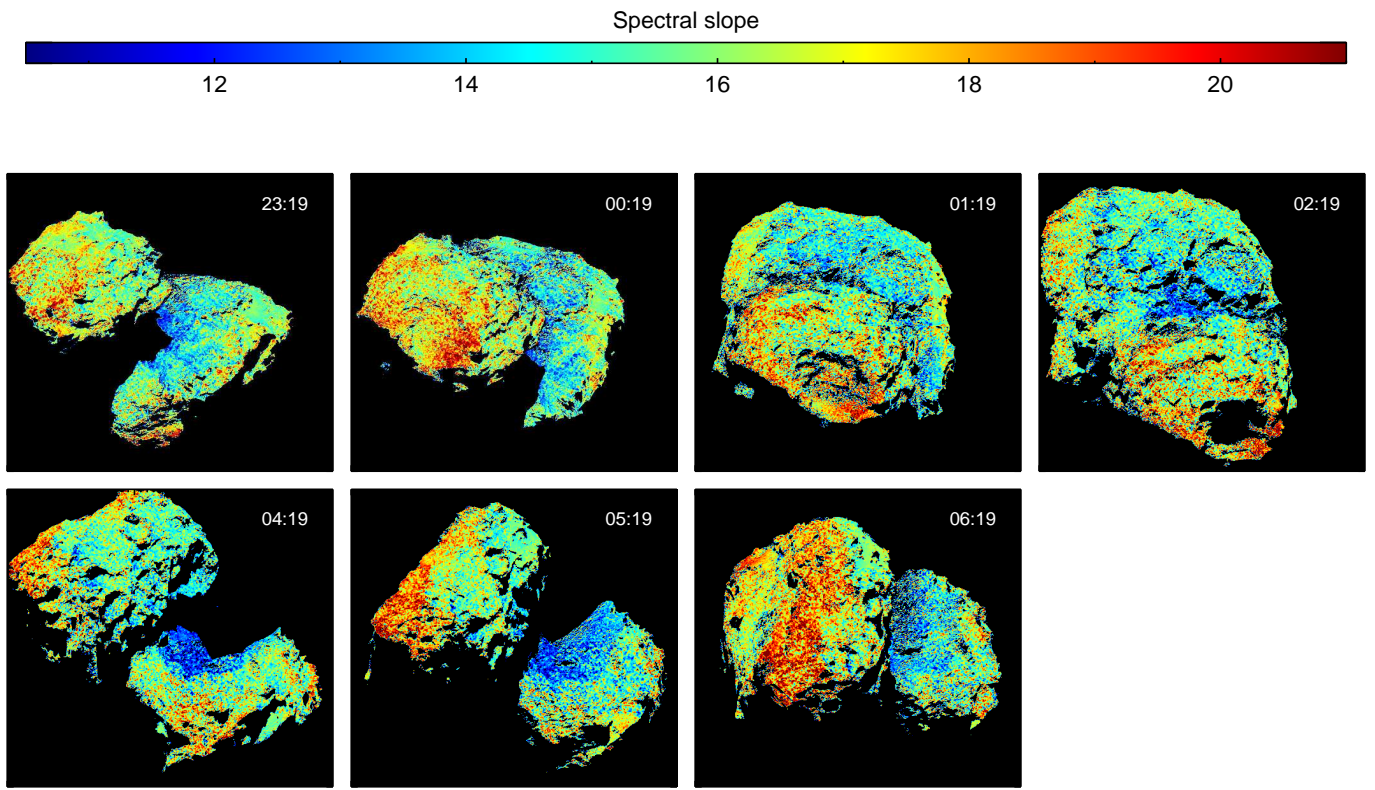}
      \caption{Spectral slope images for the observations on 6 August, at a phase angle of $\sim$ 50$^{\circ}$. The slope is computed in the 535-882 nm range, after normalization at 535 nm, and it is \%/(100 nm).}
         \label{specmap_6aug}
   \end{figure*}

According to the albedo and color variations described in the previous paragraphs, the surface of the nucleus shows some variability. To investigate the correlations between geology and spectral properties, we computed the spectral slope maps (in \%/100 nm, evaluated in the 882-535 nm range using the same method as described in Sect. 4.3) for the observations obtained on 1 and 6 August at different comet rotational phases. Figures 13 and 14 show
these spectral slope images, acquired when Rosetta was at a distance of about 800 km and 120 km from the comet, which yielded a spatial resolution of 14 m/px and 2 m/px, respectively. The spectral slope was evaluated as the mean value over a 3$\times$3 pixel squared box. These maps confirm the peculiarity of the Hapi region at increasing resolution scales. They also show the phase-reddening effect previously seen in the disk-averaged spectrophotometry, as the spectral slope of the same areas increases in the 6 August maps, when the phase angle was $\sim$ 50$^{\circ}$, compared to the 1 August images, where the phase angle was 10$^{\circ}$.  \\
Figures ~\ref{specmap_1aug} and ~\ref{specmap_6aug} show distinct spectral slope behaviors over the nucleus surface. We therefore
propose three groups of terrains based on the strength of the spectral slope:
\begin{itemize}
\item Low spectral slope, group 1. For this group, the spectral slope comprises between 11.0 and 14 \%/(100 nm) at a phase angle
of $\sim$ 50$^{\circ}$ (Fig.~\ref{specmap_6aug}), it mostly corresponds to the blue color in Figs.~\ref{specmap_1aug} and ~\ref{specmap_6aug} at different phase angles (10$^{\circ}$ and 52$^{\circ}$, respectively). For these regions, the albedo is generally higher. The regions Hathor, Hapi, and Seth are strongly correlated with this group, although not the entire Seth region is blue. We might also add portions of the Ma'at (see Fig.~\ref{specmap_1aug} at UT 17:26) and the Babi region (see Fig.~\ref{specmap_6aug} at 23:19 UT).

\item Average spectral slope, group 2. For this group, the spectral slope comprises between 14 and 18\%/(100 nm) at a phase angle
of $\sim$ 50$^{\circ}$ (Fig.~\ref{specmap_6aug}), it mostly corresponds to the cyan-green color in Figs.~\ref{specmap_1aug} and~\ref{specmap_6aug}. Several regions are entirely correlated with this group (e.g., Anuket and Serqet), and others are partially correlated (e.g., Ma'at and Ash).

\item High spectral slope, group 3. For this group, the spectral slope is above 18\%/(100 nm) at a          phase angle of $\sim$ 50$^{\circ}$ (Fig.~\ref{specmap_6aug}), it mostly corresponds to the orange and red colors in Figs.~\ref{specmap_1aug} and~\ref{specmap_6aug}. Several regions of the nucleus are associated with this group, although only the Apis region is entirely correlated with group 3. In the small lobe, the regions around the Hatmehit depression have this high spectral slope (i.e., Nut, Maftet, Ma'at, and Bastet), and a little of the inside of Hatmehit (Figure \ref{specmap_6aug} at 01:19 UT). In the large lobe, the regions around the Imhotep depression are also correlated to the strong spectral slope (i.e., Apis, Ash, and Khepry) and a little of the inside of Imhotep (Fig. \ref{specmap_6aug} at 06:19 UT).
\end{itemize}

The three groups are observed in the two portions of the nucleus, the small and large lobes. As already stated (Sierks et al., 2015), there is no obvious color variability between the two lobes to support the contact-binary hypotheses, although at the same time, it is not ruled out either since two compositionally similar objects may have collided.
Thus the color variations of the spectral slopes on the surface may be due to morphological and/or mineralogical properties. 
We explored this idea by relying on the detailed analysis of the morphological properties of each region by El-Maarry et al. (2015).
The group 1 regions are classified in various nomenclatures (i.e., brittle, dust cover, consolidated), thus not highlighting a specific morphology. This is specifically true for the Hapi and Hathor regions, whose morphology is quite different. However, it is important to note that the Seth, Ma'at, Babi, Hathor, and Hapi regions are adjacent to each other, which concentrates the location of group 1 around the interface between the two lobes. These regions also have in common that  activity has been spotted in most of them (Sierks et al., 2015, Vincent et al., 2015).
Therefore, as shown for other comet nuclei (Sunshine et al., 2006; Li et al., 2013), the spectral slope of group 1 is most likely linked to water ice mixed with refractories on the surface, rather than a specific morphology of the nucleus.\\
As previously noted, the albedo is generally correlated to the surface texture, but is not uniquely correlated to the spectral slope. In fact, smooth active regions such as Hapi and parts of Ma'at have both a very weak slope and high albedo (Figs.~\ref{albedomap} and ~\ref{specmap_6aug}). On the other side, the Imhotep region, which has a large portion of smooth material with higher albedo, did not show signs of activity during these observations, and its spectral slope belongs to groups 2 or 3.

We also immediately noted that group 3 regions are generally located opposite  the neck region. Both ends of the nucleus are characterized by very large depressions, whose origin might be different (see Auger et al. (2015) and Pajola et al. (2015) within this issue for more details). There is no evidence that these depressions are related to impacts (Pajola et al., 2015), thus the correlation of group 3 with ejecta deposits is not supported, also given the asymmetry of group 3 around these two depressions.
Nonetheless, although only the Ma'at and Ash regions are classified as "dust covered" (El-Maarry et al., 2015, Thomas et al., 2015), Maftet, Apis, and Bastet are also adjacent to the depression, and their surfaces appear rather smooth in the portions correlated with group 3. Consequently, although the dusty regions are not clearly correlated to group 3, they are connected. This again emphasizes the relationship between surface texture and albedo. \\
The original Philae landing site, Agilkia, is analyzed in detail in La Forgia et al. (2015), who reported a detailed geomorphological and spectrophotometric study of this region and of the surrounding Hatmehit area from OSIRIS images acquired at a resolution of $\sim$0.5 m/px. These regions fall in the spectral groups 2 and 3 according to our analysis. \\
Another important aspect is that the groups' spectral slope properties are not correlated to a particular morphology that may expose material from deeper inside the nucleus. In fact, the Aten depression belongs to group 2, which represents the average properties of the surface. The Nut depression, although it shows group 3 properties, is also significantly characterized by group 2 properties. 
Therefore, the spectral slope variations of the surface do not show evidence of vertical diversity in the nucleus composition, at least for the first tens of meters. However, it must be noted that sublimation lags, airfall, and surface transport complicate our interpretation, as these processes tend to homogenize the surface layer of the nucleus;  below this uppermost layer, the subsurface might be different.

\section{Conclusions}

The data from comet 67P/Churyumov-Gerasimenko  obtained from July to mid August 2014 with the OSIRIS imaging system show the nucleus with unprecedented spatial resolution. We have presented the results from the disk-averaged and disk-resolved analysis of the photometric properties and spectrophotometry of the nucleus. Our main results are the following:\\
\begin{itemize}
\item The phase function of the nucleus of 67P from disk-averaged reflectance in the 1.3$^{\circ}$--54$^{\circ}$ phase angle range shows a strong opposition effect. The fit of the phase function with the HG model gives a G parameter of -0.13, implying a steep brightness dependence on the phase angle. The absolute magnitude reduced to the Bessel V filter is $H_v(1,1,0)$ = 15.74$\pm$0.02 mag, and the 
linear slope (for $\alpha > 7^{\circ}$) is $\beta$ = 0.047$\pm$0.002 mag/$^{\circ}$, a value very similar to that found for comets Hartley 2 and Tempel 1, and close to the average value for Jupiter-family comets.
\item We presented disk-integrated spectrophotometry in 20 NAC and WAC filters covering the 250-1000 nm range. The spectral behavior of the nucleus is red, meaning that the reflectance increases with wavelength, and it is featureless in the visible and near-infrared range. This behavior is fully consistent with ground-based observations of the  comet, and it is similar to those of other bare cometary nuclei. In the mid-UV region we see an increase of the flux that indicates a potential absorption band centered on $\sim$ 290 nm, possibly due to SO$_2$ ice. However, this feature needs to be confirmed with other observations as the two WAC UV filters showing the flux enhancement are affected by pinhole defects. 
\item The analysis of the spectral slopes versus phase angle, both from disk-averaged and disk-resolved data, shows a significant phase reddening for the nucleus, with the disk-averaged spectral slope increasing from 11\%/(100 nm) to 16\%/(100 nm) in the 1.3$^{\circ}$--54$^{\circ}$ phase angle range. This behavior may be attributed to a higher contribution from multiple-particle  scattering at large phase angles and/or to surface roughness effects. The strong phase reddening suggests that the effect of multiple scattering on the 67P nucleus  may be non-negligible, even though the albedo is very low.
\item We modeled the disk-averaged reflectance in eight filters covering the 325--1000 nm range using the disk-integrated Hapke model formalism. We found no clear wavelength dependence of the $g_{\lambda}$, $B_{0}$ and $h_{s}$ parameters, implying that the shadow-hiding effect must be the main cause of the opposition surge.  The geometric albedo at different wavelengths derived from Hapke modeling perfectly matches the comet spectrophotometry behavior.
\item Hapke (2002 and 2012) modeling on disk-resolved reflectance at 649 nm gives parameters very close to those found for comets Wild 2 and Tempel 1 and is compatible with those found for comet Hartley 2, revealing that the photometric properties of these cometary nuclei are similar. The geometric albedo derived from Hapke modeling is 0.065$\pm$0.02 at 649 nm. Thus the 67P nucleus has a surface dark in absolute terms, but one of the brightest of the other cometary nuclei investigated by space missions (Li et al., 2013).
\item The nucleus shows color and albedo variations across the surface: Hapi is $\sim$ 16\% brighter than the mean albedo over the surface, while the Apis and Seth regions are about 8-10\% darker. 
\item Based on local spectrophotometry, we see an enhancement of the flux at 701 and 743 nm that is clearly due to cometary emissions from the coma located between Rosetta and the nucleus. The potential sources of these emissions are  $\rm H_2O^+$ in its (0,6,0) band at 700 nm, the (0,2,0)-(0,0,0) band at 743-745 nm, and  the band system (0,6,0) of $\rm NH_2$ , which overlaps with the water bands.
\item On the basis of the spectral slope and of the spectrophotometry, we identified three different groups of regions, characterized by a low, medium, and high spectral slope, respectively. The three groups are observed in the two lobes of the comets, and we did not note a significant color variability between the two lobes that would support the contact-binary hypothesis, although this does not exclude that the nucleus of 67P might have been formed by a collision of two objects in the past. 
\item Generally, the spectral slope values are often anticorrelated with the reflectance, the brightest regions being also the bluer ones in terms of spectral slope. Hapi, in particular, is the brightest area in absolute terms, the main source of cometary activity at large heliocentric distances, and the spectrally bluer region. We interpret this behavior as due to a larger abundance of water ice in this region, which is often cast in shadow, not illuminated at perihelion due to seasonal effects, and thus a favored region to retain ices on its surface.
\end{itemize}

\begin{acknowledgements}
OSIRIS was built by a consortium of  the Max-Planck-Institut f\"ur
Sonnensystemforschung, G\"ottingen, Germany, CISAS--University of
Padova, Italy, the Laboratoire d'Astrophysique de Marseille, France, the
Instituto de Astrof\'isica de Andalucia, CSIC, Granada, Spain, the Research and
Scientific Support Department of the European Space Agency, Noordwijk, The
Netherlands, the Instituto Nacional de T\'ecnica Aeroespacial, Madrid, Spain,
the Universidad Polit\'echnica de Madrid, Spain, the Department of Physics and
Astronomy of Uppsala University, Sweden, and the Institut  f\"ur Datentechnik
und Kommunikationsnetze der Technischen Universit\"at  Braunschweig, Germany.
The support of the national funding agencies of Germany (DLR), France (CNES),
Italy (ASI), Spain (MEC), Sweden (SNSB), and the ESA Technical Directorate is
gratefully acknowledged. We thank the referee, J.-Y. Li, for his comments and suggestions that helped to improve this manuscript.

\end{acknowledgements}

\end{document}